\documentclass[10pt, twocolumn, journal]{trans/IEEEtran}
\usepackage[utf8]{inputenc}
\usepackage{amsmath, cases, bbm, graphicx, amsthm, algpseudocode, algorithm, caption, subcaption}
\usepackage{balance} 
\expandafter\def\csname ver@subfig.sty\endcsname{}

\usepackage{xcolor, multicol, tabularx,booktabs, amssymb, dsfont, multirow, cite}

\usepackage{lipsum} % filler text
\usepackage{soul,subfig}
\usepackage[nocenter]{qtree}
\usepackage{newfloat}

\usepackage[hidelinks]{hyperref}
\usepackage{cleveref}

\usepackage{xcolor}
\usepackage{hyperref}%
\makeatletter

\newcommand\sfcodefork{%
  \ifnum\the\spacefactor=1000 \expandafter\@firstoftwo\else\expandafter\@secondoftwo\fi
}

\makeatother

\newtheorem{definition}{Definition}
\newtheorem{example}{Example}

\newtheorem{proposition}{Proposition}

\newtheorem{lemma}{Lemma}

\crefname{lemma}{Lemma}{lemmas}

\DeclareFloatingEnvironment[fileext=lod]{diagram}
\newcolumntype{C}{>{\centering\arraybackslash}X} 
\newcolumntype{s}{>{\arraybackslash\hsize=.6\columnwidth}X}

\def\figurename{Fig.}
\def\tablename{TABLE}

\begin{document}

\bstctlcite{IEEEexample:BSTcontrol}
    \title{Deadline-Aware Joint Task Scheduling and Offloading in Mobile Edge Computing Systems}
  \author{Ngoc Hung~Nguyen,
         Van--Dinh~Nguyen,~\IEEEmembership{Senior Member,~IEEE},
          Anh Tuan~Nguyen,\\
          Nguyen~Van Thieu,
          Hoang Nam~Nguyen,
          and   Symeon Chatzinotas,~\IEEEmembership{Fellow,~IEEE}
          
\thanks{N. H. Nguyen and V.-D. Nguyen are with the Center for Environmental Intelligence and College of Engineering \& Computer Science, VinUniversity, Hanoi 100000, Vietnam (e-mail: \{hung.nn,dinh.nv2\}@vinuni.edu.vn). N. H. Nguyen was also with the Department of Electrical and Electronic Engineering, Hanyang University, Ansan 15588, Korea. Corresponding author: V.-D. Nguyen.}
\thanks{A. T. Nguyen is with the Department of Smart City, Hanyang University, Ansan 15588, Korea (e-mail: natuan@hanyang.ac.kr).}%
\thanks{N. V. Thieu is with the Faculty of Computer Science, PHENIKAA University, Yen Nghia, Ha Dong, Hanoi 12116, Vietnam (e-mail: thieu.nguyenvan@phenikaa-uni.edu.vn).}%
 \thanks{H. N. Nguyen is with the Department of Information Engineering (DII), Università degli Studi di Brescia, Brescia 25121, Italy (e-mail: nam.nguyenhoang@unibs.it).}
\thanks{S. Chatzinotas is with the Interdisciplinary Centre for Security, Reliability and Trust (SnT), University of Luxembourg, L-1855 Luxembourg City,
Luxembourg (e-mail: symeon.chatzinotas@uni.lu).}
\thanks{This work is supported by the VinUniversity Seed Grant program.}
}

\maketitle
\begin{abstract}
    The demand for stringent interactive quality-of-service has intensified in both mobile edge computing (MEC) and cloud systems, driven by the imperative to improve user experiences. As a result, the processing of computation-intensive tasks in these systems necessitates adherence to specific deadlines or achieving extremely low latency. To optimize task scheduling performance, existing research has mainly focused on reducing the number of late jobs whose deadlines are not met. However, the primary challenge with these methods lies in the total search time and scheduling efficiency.
    In this paper, we present the optimal job scheduling algorithm designed to determine the optimal task order for a given set of tasks. In addition, users are enabled to make informed decisions for offloading tasks based on the information provided by servers. The details of performance analysis are provided to show its optimality and low complexity with the linearithmic time $\mathcal{O}(n\log n)$, where $n$ is the number of tasks. To tackle the uncertainty of the randomly arriving tasks, we further develop an online approach with fast outage detection that achieves rapid acceptance times with time complexity of $\mathcal{O}(n)$. Extensive numerical results are provided to demonstrate the effectiveness of the proposed algorithm in terms of the service ratio and scheduling cost.
\end{abstract}

\begin{IEEEkeywords}
Computation offloading, job scheduling, mobile edge computing, online scheduling.
\end{IEEEkeywords}
 \IEEEpeerreviewmaketitle

\section{Introduction}
\label{sec:intro}

\subsection{Background}
\IEEEPARstart{T}{he} emergence of Internet-of-Things (IoT) applications, including driverless vehicles, vehicle-to-everything (V2X), and wearable devices, underscores the increasing importance of mobile devices with low-latency and high-reliability communication. This significance is particularly pronounced in the realms of mobile edge computing (MEC), the focal point of this paper. Within this context, additional challenges arise, encompassing wireless channel conditions, hardware computation limitations, and battery constraints \cite{gul2022uav, xu2021energy, cai2021optimal}. Achieving a high quality of service (QoS) necessitates not only endowing devices with multitasking abilities to meet various completing deadlines on the application side of requests \cite{pu2018chimera} but also implementing lightweight and efficient routing protocols, along with proficient big data processing algorithms \cite{dang2020should, ji2021survey, shi2018event, cao2021resource}. A viable approach to address these requirements involves processing users' computing tasks at the edge network. This methodology is recognized as a key component in mitigating communication latency, enhancing location awareness, and minimizing the load on the core network \cite{ren2019survey}. Current studies are actively improving this method from various perspectives, including service latency \cite{gur2020convergence, porambage2018survey}, energy consumption \cite{ranaweera2021mec}, and scheduling \cite{ARUNARANI2019407, jamil2020intelligent}.

The effective management of concurrent applications, often surpassing the computational capability of devices, is a critical challenge in computing systems \cite{pu2018chimera}. Task scheduling emerges as a pivotal aspect in addressing this issue, as highlighted in recent studies \cite{HUDA2022103341}. A significant hurdle in this context is the presence of interdependent tasks, especially when cooperative systems are influenced by prolonged high-priority applications or features. Consequently, optimizing task offloading in MEC systems based on deadline constraints and task requirements becomes a paramount concern \cite{zhang2022joint, 8493149}. Moreover, achieving the dual objectives of minimizing energy consumption and total computation time for these processes is imperative to align with various constraints \cite{cai2021optimal, 9796843}. These objectives have been explored across diverse scenarios, including user mobility, UAV reconnaissance, cooperative MEC systems, and independent as well as interdependent tasks \cite{9427223, zhao2017tasks, 8486629}.

The service ratio, which is defined as the proportion of tasks served successfully before the deadline to the total number of arriving tasks \cite{zhang2010service}, is widely used to evaluate the performance of MEC systems. As this ratio approaches one, it indicates an increasing number of tasks meeting their deadlines, signifying better fulfillment of the system's requirements. Currently existing algorithms, encompassing both non-preemptive and preemptive systems, have demonstrated commendable efficiency across a substantial range of cases \cite{buttazzo2012limited}.

\subsection{Motivation and Main Contributions}
Currently, the existing scheduling solutions are inefficient and isolated in various aspects. In particular, the approximated algorithms, despite being common in the literature  \cite{ARUNARANI2019407}, often exhibit high computational cost \cite{zhou2020deep}. On the other hand, machine learning  (e.g., deep reinforcement learning \cite{10007839}) and evolutionary algorithms \cite{gong2023dependen, sriraghavendra2022dosp} can be customized to solve such problems. Additionally,  conventional algorithms can be utilized to address the offloading problems with the lower time complexity \cite{9893340, 10354444, fang2019job}.

However, the aforementioned works handle task offloading via either agent-based mobile devices or dispatching machines. This approach determines the offloading decision, which is then applied to tasks from mobile devices. However, these studies fail to address three significant problems: $i)$ Tasks may need to wait in the queue of dispatching machines without being immediately offloaded after generation; 
$ii)$ Due to the limitations of mobile device configurations, decision-making time at the inference phase may be extended \cite{rokh2023comprehensive}, or they come with the cost of the long training and looping time to arrive at the convergent point or the optimal solution \cite{zhou2020deep, ARUNARANI2019407, SHISHIDO2018378}; and $iii)$ The uncertainty of future task arrivals complicates offloading and scheduling, making these processes particularly less efficient and practical. For example, users may not be aware of the future task arrival rate and the system states of servers, resulting in a lower service ratio \cite{chen2011optimal}.
Moreover, tasks with stringent deadlines require extremely low latency on the user side, causing little room for alteration. Thus, deadline-aware task scheduling presents a promising solution to overcome these limitations. 

To address the aforementioned challenges, we introduce an optimal algorithm designed to determine the best possible task order. The primary goal of the proposed algorithm is to reduce the number of outages, thereby maximizing the system's service ratio. Additionally, we develop a fast outage detection algorithm to lower the scheduling cost in online scheduling scenarios, thus reducing the cost associated with making offloading decisions. The main contributions of this paper are summarized as follows:
\begin{enumerate}
     \item We first introduce a decentralized offloading scheme, where the agent is located on the mobile device side. This agent utilizes server information to reschedule tasks as new ones arrive. Based on the scheduling results, the agent selects the server that minimizes outages and optimizes the total completion time.
     \item We propose a novel and optimal job scheduling algorithm that aims to maximize the number of served tasks before their deadline in the non-preemptive and one-core scenario. Based on the available tasks in servers, users are empowered to select the appropriate server providing the lowest total latency. On the other hand, servers determine the optimal set of task orders based on their deadlines to maximize the overall service ratio. The proposed algorithm exhibits a low complexity by reducing from factorial to linearithmic time complexity $\mathcal{O}(n\log n)$, where $n$ is the total number of tasks. 
      \item To further reduce the computational complexity, we further develop a fast outage detection method. The key idea of the fast outage detection (FOD) method is to quickly verify whether an incoming task can be accepted for processing in an MEC server. Given the received information from servers, each user acts as an agent to decide whether to offload its new task or not. The complexity is shown to be the linear time
    $\mathcal{O}(n)$. 

    \item We provide extensive numerical results to demonstrate the effectiveness of the proposed algorithm. The results reveal the superior performance of the proposed algorithm over the state-of-the-art solutions in a wide range of settings.  
\end{enumerate}

The remainder of the paper is organized as follows. The related work is discussed in Section \ref{relatedwork}. Section \ref{sec:problem_formulation} presents the system model, task organization, and the problem formulation. The proposed algorithms are introduced in Section \ref{sec:propose_sc} with the optimal job scheduling method and the correctness proof. Section~\ref{sec:experiment} provides and discusses simulation results. Section~\ref{sec:conclus} concludes the paper.

\section{Related Work}\label{relatedwork}
In general, existing works on task scheduling and offloading mainly focus on optimizing energy consumption and reducing service latency. First, the works in \cite{8486629, 9893340, 10354444} employed conventional algorithms to address the energy consumption minimization problem. Conversely, studies in \cite{10354444, fang2019job, meng2019dedas} aimed at minimizing latency or maximizing the number of completed tasks using similar methods. The second category explored the potential of evolutionary computing and machine learning to enhance service quality under uncertain system conditions \cite{sriraghavendra2022dosp, 9151371, 9238937}.

On one hand, task scheduling plays a crucial role and is applied across various areas within computing systems. In single-core systems, the task scheduling method can generally be categorized into two primary groups: Elementary and advanced algorithms. 
In particular, the elementary algorithms developed earlier, such as rate monotonic and deadline monotonic, prioritize simplicity over adaptability. Rate monotonic scheduling (RMS) assigns priorities according to task rates, giving higher priorities to tasks with shorter periods. Moreover, deadline monotonic scheduling (DMS) organizes its representations based on earlier deadlines, using criteria such as smallest deadline, Moore--Hodgson, \text{D*S}, or least laxity first \cite{zhang2010service, fang2019job}. In essence, the RMS mechanism is akin to a first-come-first-serve sorting approach, with various versions available for scheduling or ordering tasks. In contrast, DMS is more intricate in its design, considering both deadline and processing time. The notable advantages of this category lie in the minimal scheduling cost (in terms of time) and deterministic correctness. Moreover, advanced algorithms, representing a more contemporary approach, can be further divided into machine learning-based techniques (i.e., deep reinforcement learning, imitation learning) and evolutionary computing-based methods including PSO, genetic algorithm, and ant colony optimization-based schemes \cite{9151371, QIN2005885, PAPAZACHOS20091276, ARUNARANI2019407}. However, it is important to note that these methods do not always guarantee correctness, and their high scheduling costs represent a primary drawback\cite{9151371}.

Scheduling algorithms can be further categorized into online and offline schemes, with the primary distinction lying in their awareness of arrival events within the system \cite{chen2011optimal, meng2019dedas}. The online scheme operates without complete knowledge of all incoming events, adding a layer of complexity and a closer representation of real-world scenarios. Consequently, many works delve deeper into online scheduling due to its heightened challenges and enhanced reflection of reality. For example, Wang \textit{et al.} \cite{9151371} introduced and implemented imitation learning in vehicle-enabled edge computing for minimizing latency through online job scheduling. Recently, deep learning algorithms have also been developed to address the task latency issue \cite{9145425, 9238937, ale2021delay}, that show strong applicability in dynamic systems where queuing experiences frequent changes. Some initiatives focused on reducing energy consumption through optimized task scheduling \cite{8486629, xu2021energy}. 

Regarding evolutionary computing, the works in \cite{samanta2018latency, samanta2019battle, aburukba2020scheduling} successfully tackled the challenge of minimizing end-to-end service and computation delays through a search-based approach. Unlike deep learning, evolutionary computing faces challenges when applied to dynamic systems. Despite this distinction, both methodologies share a common drawback: high scheduling costs, which pose a significant impediment to tasks with deadlines. To address these challenges, this paper introduces a novel scheduling algorithm accompanied by a correctness proof. The primary objective of this design is to be suitable for online schemes, with a specific focus on minimizing scheduling costs, particularly for tasks characterized by stringent deadlines.

\section{System Model and Problem Formulation}
\label{sec:problem_formulation}

\subsection{System Model}
\begin{figure}[!htbp]
\centering
\includegraphics[width=1.05\columnwidth]{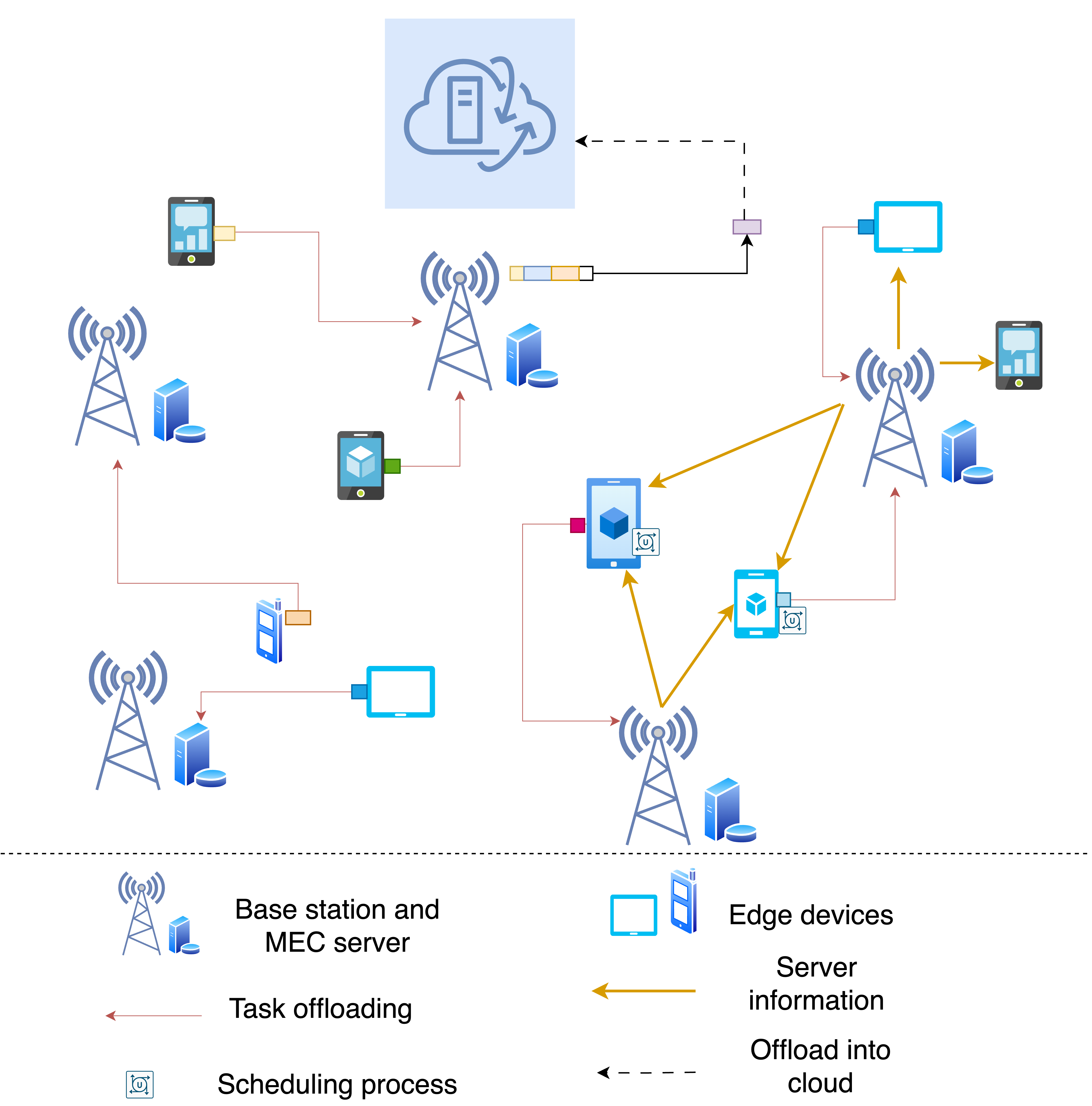}
\caption{Illustration of the considered MEC system with distributed edge servers.}
\label{fig:systemmodel}
\hfill
\end{figure}

We consider a MEC system as illustrated in \autoref{fig:systemmodel}, consisting of $E$ distributed edge servers and $K$ users in a network area $\mathcal{A}$. The MEC servers are represented by the set $\mathcal{S} = \{S_1, S_2, \cdots, S_E\}$, where each is co-located with a base station (BS). The position of server $S_e$ is denoted by  $\boldsymbol{a}_e\in \mathcal{A}$, $e\in [1,E]$.  We assume that the orthogonal bandwidth $W/E$ [Hz] is assigned to each associated BS so that $N$ independent communication channels are available to each server. 
By the frequency division multiple access (FDMA), the channel bandwidth allocated to each channel is $\frac{W}{EN}$.

We assume each user generates multi-computation tasks or jobs that can be offloaded to one server for processing. Each task is independent of the others and has a limited processing time. We model the generation of tasks by the $k$-th user equipment (UE) as following a Poisson distribution with an arrival rate of $\lambda_k$. Consequently, at any specific time, only one task is generated in the system and is offloaded immediately. Additionally, we assume that the server can instantly detect changes in its queue as soon as a user selects it for task offloading. Let $t^{[\textsf{a}]}_{i,k}$ denote the arrival time of task $i$. The tuple of task $i$ is defined as by $ T_{i,k}\triangleq(t^{[\textsf{a}]}_{i, k}, \alpha_{i,k}, \beta_{i,k}, \mathsf{Z}_{i,k})$, where $\alpha_{i,k}$ is the number of CPU cycles required for $T_{i,k}$, $\beta_{i,k}$  is the deadline for completion, and $\mathsf{Z}_{i,k}$ is the data size of task required for transmitting from UE $k$ to the associated MEC. We further assume the task is atomic or indivisible and is handled by a single thread. A multi-core server can simultaneously handle multiple tasks by assigning each task to a separate core. For simplicity, we consider that each server has only one--core, and preemptive is not considered in this work. Therefore, each server can process only one task at a time.

Suppose that the position of UE $k$ is denoted by $\boldsymbol{a}^{[\textsf{u}]}_k(t)\in \mathcal{A}$. For simplicity, we consider both BSs and users to be equipped with a single antenna. The channel coefficient from user $k\in\mathcal{K}\triangleq[1,\cdots,K]$ to channel $n \in\mathcal{N} \triangleq[1,\cdots,N]$ of server $S_e\in\mathcal{S}$ at time $t$ is modeled as
\begin{align}
h_{k,e,n}(t)=\frac{g_{k,e,n}(t)}{\big\|\boldsymbol{a}_e-\boldsymbol{a}^{[\textsf{u}]}_k(t)\big\|^{p/2}}.
\end{align}
Herein, $p\in[2,\, 4]$ is the path-loss exponent and $g_{k,e,n}(t) \sim \mathcal{CN}(0,1) $ denotes the small-scaling fading. We assume that the channel is modeled as a flat fading channel, whose gain is constant during one time slot, but changes independently from one time slot to the next. By FDMA, the achievable rate of UE $k$ using channel $n$ connected to server $m$ at time $t$ can be expressed as 
\begin{align} \label{eq:uploading_rate}
R_{k,e,n}(t)= W_{c}\log\left(1+\frac{|h_{k,e,n}(t)|^2P_k}{W_c N_0}\right)
\end{align}
where $W_c = \frac{W}{NE}$ denotes the channel bandwidth, $P_k$ [Watts] is the transmit power of UE  and $N_0$ [Watts/Hz] is the noise power spectral density.

%\subsection{MEC Strategy}

\subsection{Offloading Model}
\label{se:off}
Users need to determine the most appropriate server to offload tasks. We assume that the number of servers within the network coverage is $L$. We denote $\mathcal{S}_k \subset \mathcal{S}$ as the set of all nearest servers with UE $k$, such as $|\mathcal{S}_k| = L$. The user can offload its tasks to the appropriate  $L$ servers.
\begin{figure}[t]
\centering
\includegraphics[width=1.1\columnwidth]{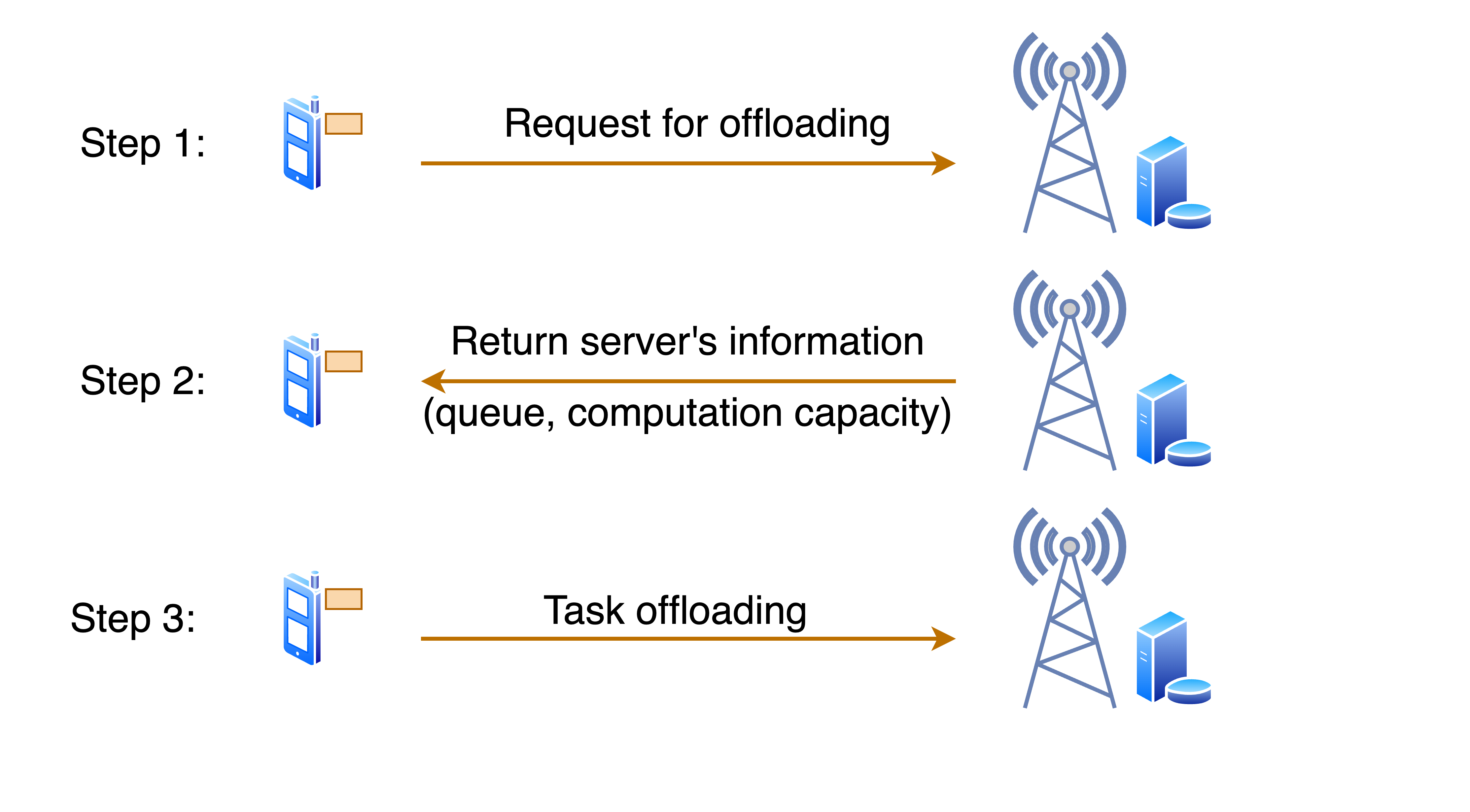}
\caption{The three-step offloading process.}
\label{fig:offloadingsteps}
\hfill
\end{figure}

\autoref{fig:offloadingsteps} illustrates the process to determine the offloaded MEC from the user side, which is detailed as follows:
\begin{enumerate}
    \item During an arrival event,  users reach out to nearby servers to update task details such as input data, CPU cycle demands, and deadline requirements.
    \item Upon receiving the updated information, servers return their status, encompassing details like remaining tasks in queues and computational capacity. It is noteworthy that the remaining tasks include all tasks designated for offloading to the server, regardless of whether they arrive after the current task.
    \item In the final step, the user offloads the actual tasks to the selected server. A server is chosen based on its offloading capacity, as determined by the provided information. In this context, the user selects a server if it offers the minimum outage and total computation time, provided there is at least one available channel for communication.
\end{enumerate}
The server operates within a dynamic queue framework, wherein updates to the server's queue occur after an arrival event at the server. As a result, the task order within the queue may change, potentially leading to tasks experiencing outage in the worst-case scenario due to this update. Hence, minimizing any delay before the task's processing is crucial.

\begin{figure}[t]
\centering
\includegraphics[width=0.9\columnwidth]{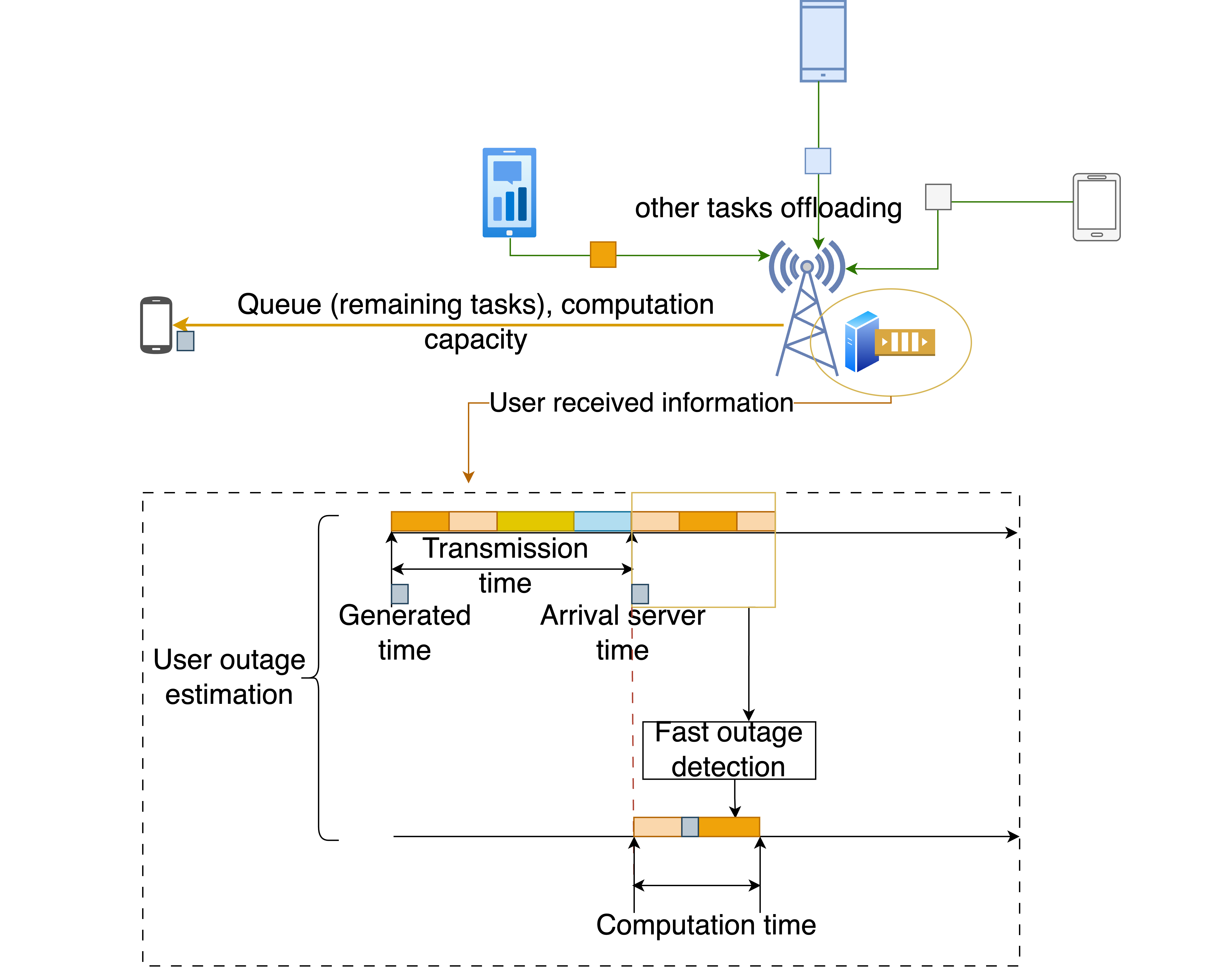}
\caption{Users estimate outages based on the servers' status.}
\label{fig:Serverinformation}
\hfill
\end{figure}

As depicted in \autoref{fig:Serverinformation}, the user gathers information from all nearby servers, assesses outage estimates, and chooses one for offloading. The critical sub-step here is scheduling, which must meet two conditions: correctness and short scheduling time.
It is important to note that the latency involved in transmitting results back to the user is relatively negligible compared to the delays caused by transmission and computation within the server. Hence, in this paper, we disregard the transmission delay when sending data back to users.

\subsection{Problem Formulation}
We define the initial position of UE $k$ as $\boldsymbol{a}_k(0)$, which is assumed to be randomly and uniformly distributed over the network area. At  time $t$, the position is determined by
\begin{align}
    \boldsymbol{a}^{[\textsf{u}]}_k(t) = \boldsymbol{a}^{[\textsf{u}]}_k(t-\Delta) + \delta_k\mathbf{d}^{[\textsf{u}]}_k(t)
\end{align}
where $\delta_k$ is the traveled distance,  $\mathbf{d}_k(t) = [\cos{\omega_k(t)},\ \sin{\omega_k(t)}]$ is the direction vector that  UE $k$ have changed in time interval $\Delta$. For the direction angle, we model it as a Markov random with the following:
\begin{align}
    \omega_k(t) = 
    \begin{cases}
        \omega_k(t-1) \text{ with probability } 1-p;\\
        \parbox[t]{0.5\columnwidth}{Random value in the range of $[0:2\pi]$ with probability $p$} 
    \end{cases}
\end{align}
where $p = p_0\textbf{f}(1/(\Delta))$ is a direction probability. This is described as the function of time, with $p_0$ being the given probability. At the beginning, we assume all users go straight, leading to the initial value $\omega_1 = 0$. We assume that UE $k$ moves at the constant speed $v_k$, \textit{i.e.} $\delta_t = \Delta v_k$

\subsubsection{Transmission Modeling}

Let $N_e^{\text{avi}}$ be the number of available channels in the associated base station with server $S_e$. We assume that each task is scheduled to offload immediately after it is generated. Therefore, there is no waiting time in queuing for transmission. Given the task size $\mathsf{Z}_{i,k}$, the transmission delay  for task $T_i$ at time $t$ of UE $k$ via channel $n$ to MEC $S_e$ can be expressed as
\begin{align}
    \tau_{i,k,e}^{\text{[trans]}} = \frac{\mathsf{Z}_{i,k}}{R_{k,e,n}(t)}.
\end{align}

\subsubsection{Computational Modeling}
During an inter-arrival time at the server, the queue and its tasks are processed as the order provided by \autoref{alg:1}.
The task's computing delay is determined by the waiting time in the queue and the execution time. In this paper, the queue is dynamic, and the order may be changed in response to an arrival event in the server. That implies that if a task is not processed between two continuous arrival events either its order in the queue will be changed, or it becomes outage after a new arrival event in the server.

We define $\mathbf{s}_e= \{s_e(1), s_e(2), \cdots, s_e(q_m)\}$ as the current order in the server $S_e$'s queue, where $q_e$ is the number of tasks in the queue. Each $s_e(l)$ with $l\in [1, q_e]$ is a pair of integers $(i,k)$ which indicates task number $i$-th at UE $k$. The computation delay $\tau_{s_e(l),e}^{\text{[comp]}}$ can be expressed as
\begin{align}
    \label{eq:tot_latency}
    \tau_{s_e(l),e}^{\text{[comp]}} =  \frac{1}{f_e}\sum_{j=1}^l\alpha_{s_e(j)}
\end{align}
where  $f_e$ is the computation capacity of the MEC server $S_e$.
\begin{definition}
    A task is considered to be outage when its overall latency is larger than the completion deadline, \textit{i.e.}
    \begin{equation}
    t_{i,k,e} = t^{[\textsf{a}]}_{i, k} + \tau_{i,k,e}^{\text{[trans]}} + \tau_{s_e(l),e}^{\text{[comp]}} > \beta_{i,k}.
    \end{equation}
\end{definition}

Let $G^{e}$ be the total generated events in the system. It is obvious that, if the system is considered in a long-term setting, $G^{e}$ will reach infinity. 
Now, we describe the decision for the $G^{e}$ arrival by the matrix $\mathbf{D}^{G^{e}\times E}$,  where each row describes the user decision at a time instant. For the row $i$, if server $S_j$ is selected for offloading, we have $\mathbf{D}_{i,j} = 1$; Otherwise, $\mathbf{D}_{i,j} = 0$. Note that, in this paper, the task cannot be divided into parts, and it will be offloaded to one server, \textit{i.e.} $\sum_{j=1}^E \mathbf{D}_{i, j}=1$.
We further consider $\mathcal{D}$ as all the possible decision matrices for $G^{e}$ generated events. Aiming to minimize the number of outages, we consider the following optimization problem
\begin{subequations} \label{eq:JFCS1}
	\begin{IEEEeqnarray}{cl}
		& \min_{\mathbf{D}\,\in \mathcal{D}} \quad
  \sum_{i=1}^{G^{e}} \sum_{j=1}^E\mathbbm{1}(\mathbf{D}_{i,j}=1)\sum_{l=1}^{q_j} \mathbbm{1}\big( t_{s_j(l),j} > \beta_{s_j(l)}\big)
    \label{eq:min_p}\\
&\text{Subject to} 
\quad  \sum_{j=1}^E \mathbf{D}_{i,j}=1,\ \forall i \in [1, G^{e}],
\label{eq:cons1}\\
&\qquad\qquad\quad\sum_{j=1}^E \big( \mathbbm{1}(D_{i,j}=1) N_j^{\text{avi}} \big)> 0,\
\forall i \in [1, G^{e}]. \label{eq:JFCS1e} \quad
	\end{IEEEeqnarray}
\end{subequations}
In the objective function \eqref{eq:min_p}, the first indicator shows whether the arrival event number $i$-th offloads to server $S_j$ or not. The second indicator function represents an outage event. Constraint \eqref{eq:cons1} specifies that only one selected server, while \eqref{eq:JFCS1e} indicates that the selected server must have available communication channels.
 
\begin{proposition}
    The problem described in \eqref{eq:JFCS1} is  NP--hard.
\end{proposition}

\begin{proof} Let us start by considering a special case. Given a MEC server with tasks arriving randomly, each task is assigned a specific deadline. Problem \eqref{eq:JFCS1} is reduced to
    \begin{subequations} \label{eq:NPhard}
	\begin{IEEEeqnarray}{cl}
		& \min \quad
        \sum_{i=1}^{G^{e}} \sum_{l=1}^{q_j} \mathbbm{1}\big( t_{s_j(l),j} > \beta_{s_j(l)}\big)
            \\
        &\text{Subject to } 
         N^{\text{avi}}_j > 0.  \quad
        \end{IEEEeqnarray}
    \end{subequations}
   Problem \eqref{eq:NPhard} is similar to $1|r_i|\sum_i U_i$ in the scheduling problem, where $1$ stands for a mono-core system, $r_i$ is the arrival time for task $i$  being executed in the system, and $U_i$ denotes an indicator function that points out whether the task is completed on time. By \cite{lawler1994knapsack}, this special case is shown to be NP-hard, and thus, so is the original problem \eqref{eq:JFCS1}.
\end{proof}

We note that solving \eqref{eq:JFCS1}, which is an NP-hard problem, optimally is very challenging, if not impossible. The main challenge of solving problem \eqref{eq:JFCS1}: It arises not only from random arrival events but also by searching for solutions for a specific set of tasks. In particular, given a set of $n$ tasks, the basic idea is to find the whole possible orders with maximum served tasks. This will take  $n!$ times for searching the solution. Moreover, the random tasks may require to be processed on time subject to diverse and stringent deadlines. Thus, the scheduling cost and scheduling correctness are crucial to maximize the service ratio. In the following section, we develop the optimal job scheduling algorithm with low complexity for solving it effectively.

\section{Proposed Algorithms}
\label{sec:propose_sc}
Ultra-low communication latency requirements necessitate that each step of the offloading process is executed within an extremely short timeframe. To achieve this, we propose two key approaches. Firstly, we introduce an optimal job scheduling method aimed at maximizing the number of tasks completed before their predefined deadlines. Secondly, we endeavor to devise an online algorithm capable of swiftly identifying whether an incoming task on servers constitutes an outage, and determining the requisite waiting time for task execution on each server.

\subsection{The Optimal Job Scheduling}
The key idea behind optimal job scheduling is composed of two steps: 1) determine a set of the order based on given deadlines, and 2) select the task with minimum data size and place it into the corresponding position. Unlike the work in \cite{CHERIYAN2021842}, if the selected task violates the completion deadline, this task will be rejected; Otherwise, it will be admitted in a queue without deletion. Details are described as follows.\footnote{In this part, tasks are considered to be arriving at the same time. Thus, the generated time is not mentioned in the whole subsection.}

Let us define $\mathbf{c}\triangleq[c(1), c(2), \cdots, c(N)]$ as the set of required CPU cycles ordered $[1:N]$ satisfying  $\alpha_{a(i)}\leq\alpha_{a(j)}$ for all $i,j\in[1:N]$ with $i<j$. 
If $\alpha_{a(i)}=\alpha_{a(j)}$, then $\beta_{a(i)}\geq\beta_{a(j)}$ is true for all $i,j\in[1:N]$ with $i<j$. 
Similarly, denote by $\mathbf{b}\triangleq[b(1), b(2), \cdots, b(N)]$ the order set of $[1:N]$ satisfying $\beta_{b(i)}\leq\beta_{b(j)}$ for all $i,j\in[1:N]$. 
Note that if $\beta_{b(i)}=\beta_{b(j)}$,  $\alpha_{b(i)}\leq\alpha_{b(j)}$ holds for all $i,j\in[1:N]$ with $i<j$. 
The proposed optimal job scheduling algorithm is summarized in \autoref{alg:1}. To make it clear, we provide the following example.
\begin{algorithm}[t]
\caption{Optimal Job Scheduling}
\label{alg:1}

\begin{algorithmic}[1]

\State{{\bf Initialization}: Set $\mathbf{q} =  [q(1), q(2), \cdots, q(N)]= \mathbf{0}_{1\times N}$.}

\For {$i\in [1:N]$}
    \State {\bf Matching required CPU cycles}:
    \State {\quad Find index $i^*$ such that $b(i^*)=c(i)$.}
    \State {\quad Update $q(i^*) \gets b(i^*)$.}
    \State {{\bf Outage check}: Set outage $= 0$.}
    \For{$j\in [i^*:N]$}
        \State{Define $\alpha_0 = 0 $}.
        \If{$\left(q(j)\neq 0\right)$ and $\left( \sum_{k = 1}^{j}\frac{\alpha_{q(k)}}{f}>\beta_{q(j)} \right)$}
        \State {Update outage $\gets$ 1.}
        \EndIf
        \If{outage $==$ 1}
        \State{Update $q(i^*)$ $\gets$ 0.}
        \State{\textbf{break}}
        \EndIf
    \EndFor
\EndFor

\State{{\bf Return}: Scheduling set $\mathbf{q}\backslash\{0\}$.}
\end{algorithmic}
\end{algorithm}

\begin{table}[htbp]
\caption{List of Tasks}
\label{tab:0}
\begin{tabularx}{\columnwidth}{@{} >{\color{black}}p{0.075\columnwidth}| >{\color{black}}C >{\color{black}}C >{\color{black}}C @{}}
\toprule
ID  
 & Description & Required Cycles [Mcycles]  & Deadline [ms] \\ 
\midrule
1 & Facial Recognition &4 &8 \\
2 & Authorization Verification &5 &6 \\
3 & Intelligent Car Signal&2 &11 \\
4 & GPS Synchronization &1 &6 \\
5 & Traffic Analysis &2 &4\\
\bottomrule
\end{tabularx}
\end{table}

\begin{example}
    \label{ex:1}
     In practice, users may offload various tasks to the MEC system to minimize latency, particularly for applications related to emergency and safety, as well as autonomy and driving enhancement. Therefore, we focus on use cases for the MEC system that assist intelligent transportation systems. Tasks such as facial recognition, authorization certification, and GPS synchronization are essential for the operation of vehicles in these systems. Typically, latency must be under 40 ms when tasks are handled by a cloud server and under 5 ms when processed by MEC servers \cite{arthurs2021taxonomy}.
    
   As shown in \autoref{tab:0}, five tasks with IDs of $[1:5]$. This example explains how \autoref{alg:1} schedules these tasks. For this scenario, the server's CPU frequency is set to $f=1$ [Bcycle/s].
    Under the ordered sets with there required CPU cycles $\mathbf{c}$ and deadline $\mathbf{b}$, we obtain $\mathbf{c} = [4,5,3,1,2]$, $\mathbf{b} = [5,4,2,1,3]$. At the beginning, the queue set $\mathbf{q} = [0,0,0,0,0]$ is initialized corresponding to five tasks in \autoref{tab:0}. At the scheduled time, this set will be filled by task IDs.
    For the first selection, \autoref{alg:1} selects the task with the minimum number of required cycles (\textit{i.e.} $c(4)=1$. It is matched with Task 4 corresponding to index 2 in set $\mathbf{b}$. Then \autoref{alg:1} updates $q(2) = 4$ and $\mathbf{q} = [0,4,0,0,0]$. As the final step, \autoref{alg:1} checks whether Task 4 violates the deadline condition or not. It is clear from \autoref{tab:0} that Task 4 satisfies the deadline condition.

     Similarly, repeating the same steps with Tasks 4, 3, and 1 leads to an update $\mathbf{q} = [5,4,0,1,3]$. With Task 2, the updated $\mathbf{q} = [5,4,2,1,3]$ raises a violation of the deadline condition of Task 2. Therefore, \autoref{alg:1} removes this ID from set $\mathbf{q}$ by updating $q(3)=0$ and $\mathbf{q} = [5,4,0,1,3]$.
As a result, the optimal order produced by \autoref{alg:1} will be $\mathbf{s}' =\mathbf{q}\backslash\{0\} = [5,4,1,3]$.
\end{example}

In \autoref{alg:1}, the certain elements of $\mathbf{q}$ may be equal to zeros because some tasks are not scheduled, leading to their removal from the final results. Consequently, the actual size of scheduled tasks in $\mathbf{q}$ may not be $N$. To address this issue, we introduce the new set $\mathbf{s}' = [s'(1), s'(2), \cdots, s'(P)]$, which exclusively contains non-zero elements, with a size denoted as $P \in [0:N]$. Here, $P=0$ indicates that no tasks are scheduled, while $P=N$ signifies that all tasks are scheduled by \autoref{alg:1}. The subsequent section will provide a comprehensive performance analysis and mathematical proof.

\subsection{Time Complexity Analysis and Optimality Proof}

\subsubsection{Time Complexity Analysis of \autoref{alg:1}}
We recall that \autoref{alg:1} include the two main steps:
\begin{itemize}
    \item Traversing all elements in set $\boldsymbol{c}$ costs $N$ times.
    \item In each time of traversing an element in $\boldsymbol{c}$, \autoref{alg:1} verifies outage by checking all elements in set $\mathbf{q}$ it spends $N$ times. This work corresponds with the finding of an element in set $\mathbf{q}$ with a minimum value of $\sum_{k=1}^j \frac{\alpha_k}{f}-\beta_j$. An outage occurred if this value is smaller than 0, resulting in the complexity of $\mathcal{O}(\log N)$.
\end{itemize}
Thus, the time complexity of \autoref{alg:1} is $\mathcal{O}(N^2)$, and that of the searching-based algorithm is $\mathcal{O}(N\log N)$.

\subsubsection{Optimality Proof}
In this work, the replacement argument method \cite[Chapter~13]{roughgarden2022algorithms} is applied to demonstrate the correctness of the proposed algorithm.

\begin{lemma}
\label{lem:1}
Assume that $\mathbf{s}\triangleq[s(1),s(2),\cdots,s(S)]$ is a scheduling set without outage. We denote $\mathbf{s}'=[s'(1),s'(2),\cdots,s'(S)]$ as the scheduling set constructed from $\mathbf{s}$ such that $\beta_{s'(i)}< \beta_{s'(j)}$ for all $i, j= i+1$, where $i,j\in[1:S]$. 
Then, the scheduling set $\mathbf{s}'$ also has no outage.
\end{lemma}
\begin{proof}
See Appendix \ref{p:lamma1}.
\end{proof}
In \cref{lem:1}, we demonstrate that after sorting by increasing the task deadlines, the number of outages is unchanged. Thus, it is possible to choose an optimal set as an increasing order of deadlines among non-unique optimal solutions. Next, we provide insights into the position and the summation of tasks in comparison with an optimal set, which are presented in Lemmas \ref{lem:2} and \ref{lem:3}. 
\begin{lemma}
\label{lem:2}
For the optimal ordered set $\mathbf{s}$ without outage and the ordered set $\mathbf{s}'$, if there exists an index $i\in[1, P]$ such that tasks $T_{s(i)}$ and $T_{s'(i)}$ satisfy $\alpha_{s(i)} < \alpha_{s'(i)}$ and $\beta_{s(i)} > \beta_{s'(i)}$, and all elements in $[1:i-1]$ are indentical in both $\mathbf{s}$ and $\mathbf{s}'$, we have  $s(i)\in \mathbf{s}'$. 
\end{lemma}

\begin{proof}   
 Based on the selection steps in \autoref{alg:1}, $s(i)$ is selected before the selection of $s'(i)$. On the other hand, if $\beta_{s(i)} > \beta_{s'(i)}$,  $s(i)$ is positioned behind $s'(i)$ in $\mathbf{s}$.
As a result, the inequality $\sum_{k=1}^{i}\alpha_{s'(i)} \leq \beta_{s'(i)} \leq \beta_{s'(l)}$ holds true. This indicates that $s(i)$ belongs to set $\mathbf{s}$ with index $i$ due to the order of selection. However, given the present of $s'(i)$ in  $\mathbf{s}'$ and $\beta_{s(i)} > \beta_{s'(i)}$,  the index of $s(i)$ in $\mathbf{s}'$ is larger than $i$.
\end{proof}

\begin{lemma}
\label{lem:3}
     For the index $i\in [1:P]$ such that tasks in the range of $[1:i-1]$ are the same in both sets $\mathbf{s}$ and $\mathbf{s}'$. If there exists an index $j \in [i+1:P]$ with $\beta_{s(i)} < \beta_{s'(i)}$, $\alpha_{s(i)} \geq \alpha_{s'(i)}$, and $\beta_{s(j)} \leq \beta_{s'(j)}$, then   $\sum_{l=i}^j\alpha_{s(l)} \geq \sum_{l=i}^j\alpha_{s'(l)}$ with $l \in [i:j]$.
\end{lemma}

\begin{proof}
See Appendix \ref{p:lamma3}.
\end{proof}

\begin{lemma}
\label{lem:4}
    Consider the ordered set $\mathbf{s}$ without any outages as an optimal ordered set with the size $M$, \autoref{alg:1} produces an ordered set $\mathbf{s}'$ without any outages with the size $P$. Then $\mathbf{s}'$ is an optimal ordered set and $P=M$. 
\end{lemma}

\begin{figure}
     \centering
     \begin{subfigure}[b]{1.0\columnwidth}
         \includegraphics[width=\textwidth]{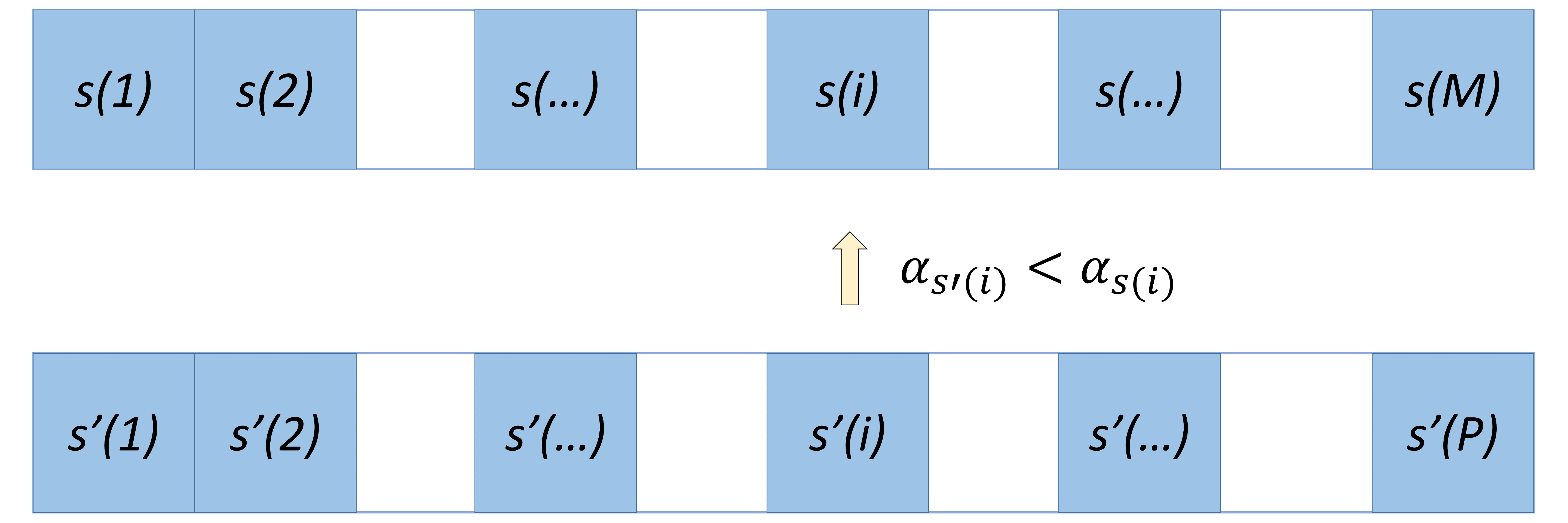}
         \caption{Cases 1 and 2.}
         \label{fig:c12}
     \end{subfigure}
     \hfill
     \begin{subfigure}[b]{1.0\columnwidth}
         \includegraphics[width=\textwidth]{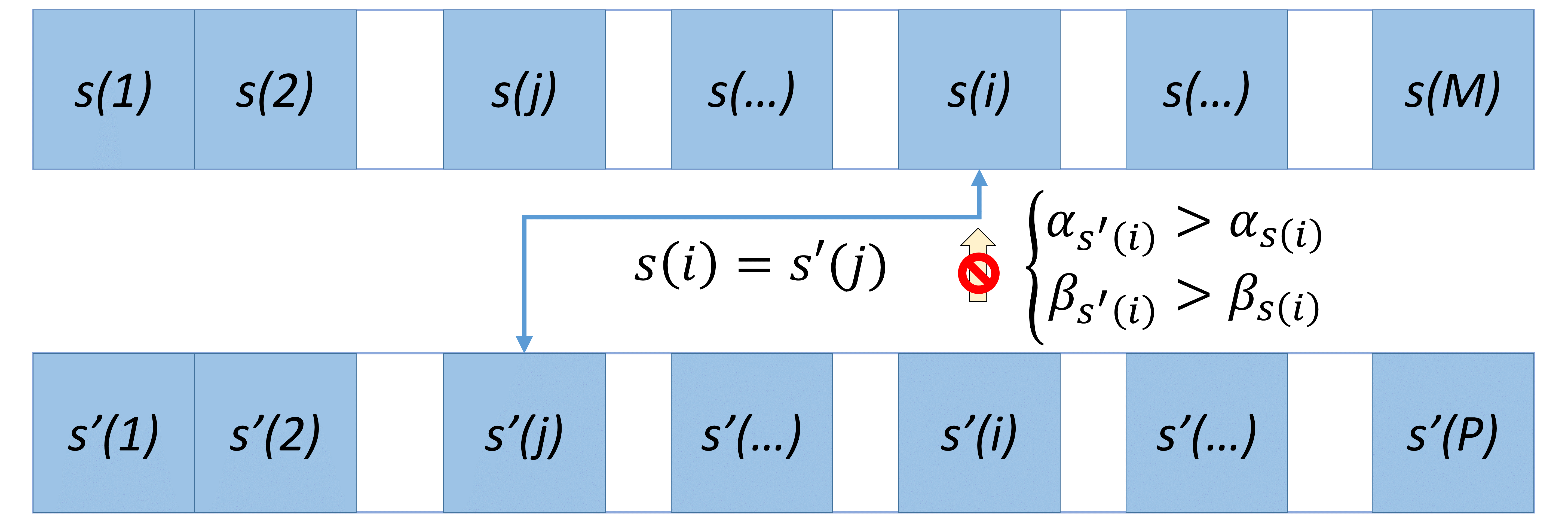}
         \caption{Case 3.}
        \label{fig:c3}
     \end{subfigure}
     \hfill
     \begin{subfigure}[b]{1.0\columnwidth}
         \includegraphics[width=\textwidth]{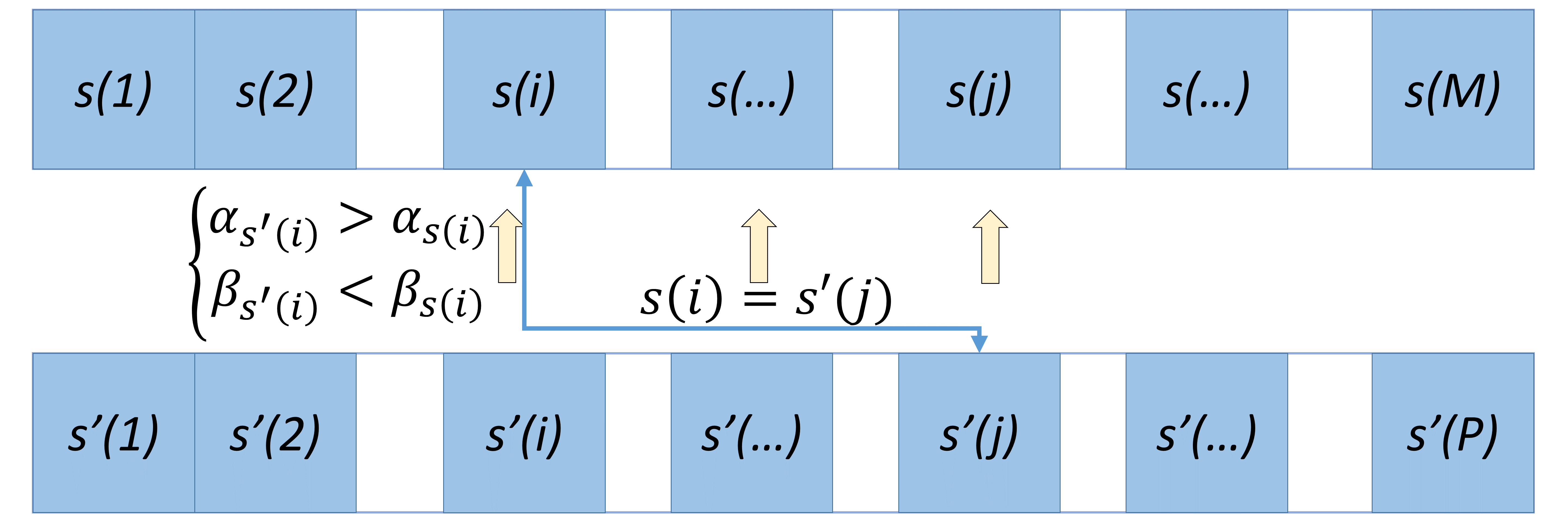}
         \caption{Case 4.}
        \label{fig:c4}
     \end{subfigure}
        \caption{The replacement cases of the proposed algorithm with the optimal set.}
        \label{fig:prs}
\end{figure}

\begin{proof}
     As mentioned previously, we apply the replacement argument method for the greedy algorithm to prove the optimality of \autoref{alg:1}. The replacement procedure is illustrated as in \autoref{fig:prs}. To proceed further, we make the following assumption.

    \noindent\textit{\textbf{Assumption 1}:} \textit{Without loss of generality, we assume that all elements in $[1:i-1]$ for any $i \in [1:M]$ in $\mathbf{s}$ are the same as in $\mathbf{s}'$. Therefore, the ordered set $\mathbf{s}' =[s'(1), s'(2), \cdots, s'(i-1), s'(i),\cdots ,s'(P)]$ is produced by \autoref{alg:1}; and the ordered set $\mathbf{s} =[s'(1), s'(2), \cdots, s'(i-1), s(i),\cdots ,s(M)]$ is an optimal set. The optimal set $\mathbf{s}$ satisfies the condition $\beta_{s(k)} < \beta_{s(l)}$ for any $k, l \in [1:M]$ and $k<l$.}

    Assumption 1 is introduced to ensure generality by suggesting that the replacement occurs at position $i$. This encompasses all preceding elements in the range from 0 to $i-1$ that have undergone replacements in the optimal set. For instance, if $i = 1$, it indicates that no replacements have occurred. Conversely, if $i = 5$, it means that all elements within the range 1 to 4 in both sets have been replaced, resulting in identical elements at this point.

    We now consider the four possible cases for the replacement. In each case, the main aim is to demonstrate that there are 1) no outages, 2) no duplication, and 3) increasing the order of deadlines in the    
    optimal set.
    \begin{enumerate}
        \item Two tasks related to $s'(i)$ and $s(i)$ satisfy $\alpha_{s'(i)} < \alpha_{s(i)}$, $\beta_{s(i)} \geq \beta_{s'(i)}$. In this case, the replacement is indeed no violation of the above conditions.
        \item Two tasks related to $s'(i)$ and $s(i)$ satisfy $\alpha_{s'(i)} < \alpha_{s(i)}$ and $\beta_{s(i)} < \beta_{s'(i)}$. In this case, the direct replacement cannot be conducted due to deadline conditions. Therefore, we develop  \autoref{alg:2} for easy-to-understand replacement steps. Combining with \cref{lem:3}, we show that the replacement satisfies the predefined constraints.
        \item The data sizes and deadlines of tasks related to $s'(i)$ and $s(i)$ satisfy $\alpha_{s'(i)} > \alpha_{s(i)}$ and $\beta_{s'(i)} < \beta_{s(i)}$. This case does not exist since it conflicts with the general assumption.
        \item The data sizes and  deadlines of tasks related to $s'(i)$ and $s(i)$ satisfy $\alpha_{s'(i)} > \alpha_{s(i)}$ and $\beta_{s'(i)} > \beta_{s(i)}$. In this case, we prove that $s(i)$ is an element in $\mathbf{s}'$ (\textit{i.e.} $s(i) \in \mathbf{s}'$). 
    \end{enumerate}
     After showing that all elements in $\mathbf{s}'$ can replace the corresponding positions in set $\mathbf{s}$, we will prove that the set of remaining elements in $\mathbf{s}$, which are not replaced, is empty.
     It should be noted that the ordered set $\mathbf{s}$ is considered as the optimal and increasing deadline set. The proof of Lemma \ref{lem:4} is detailed in Appendix \ref{p:lamma4}, where we apply Lemmas \ref{lem:2} and \ref{lem:3} to prove that these conditions are guaranteed over the replacements.
\end{proof}

\subsection{Fast Outage Detection}

\begin{algorithm}[t]
\caption{Fast Outage Detection (FOD)}
\label{alg:3}

\begin{algorithmic}[1]

\State{{\bf Input}: Optimal order set $\mathbf{s}$, task $T_a$, computation capacity of considered server $f$.}
\State{Set delay $= \text{current\_clock}$.}
\State{Construct set $\mathbf{b}$ based on tasks in $\mathbf{s} \cup a$, $a$ is the subscript of task $T_a$}
\For{$i \in [1, M+1]$}
    \If{$\alpha_{b(i)} \leq \alpha_a$}
        \State{delay $\gets$ delay + $\frac{\alpha_a}{f}$.}
        \If {delay $> \beta_{b(i)}$}
            \State{Fail to schedule $T_a$.}
            \State{\textbf{break}}
        \EndIf
    \EndIf
\EndFor
\end{algorithmic}
\end{algorithm}

The offloading process involves verifying whether an incoming task can be accepted for processing in an MEC server. This verification becomes particularly crucial as the number of nearest servers and users in the system increases, essentially scaling up the demands on the system. While the user serves as an agent in deciding where a task should be offloaded, the limitation of computational resources stands as a key obstacle. It is noteworthy that the scheduling cost directly impacts both the order and completion time of tasks. Consequently, the ability of users to rapidly detect whether offloading is necessary becomes paramount. In this subsection, we propose an online scheduling scheme aimed at improving the scheduling process with a time complexity of $\mathcal{O}(N)$. The FOD algorithm is detailed in \autoref{alg:3}.

The main idea relies on information received from nearby servers to decide whether a task should be offloaded or not. To simplify, we denote a new incoming task as $T_a \triangleq (t_{a}^{\textsf{[a]}}, \alpha_a, \beta_{a}, \mathsf{Z}_{a})$, and define $\mathbf{s} = [s(1), s(2), \cdots, s(P)]$ as the current schedule. Note that, $a$ is represented for a pair of $\{i,k\}$. To assess whether $T_a$ can be integrated into the existing order $\mathbf{s}$, the algorithm executes two primary steps: 1) forming set $\boldsymbol{b}$ with task $T_a$, and 2) filtering out tasks with CPU cycle requirements exceeding $\alpha_a$. Acceptance of $T_a$ hinges on its ability to integrate seamlessly without causing disruptions among the remaining tasks.

\subsubsection{Time complexity of FOD}
It is worth noting that applying an optimal algorithm, such as Moore-Hodgson or \autoref{alg:1}, directly to detect the outage of the new arrival task at the user side would result in a time complexity of $\mathcal{O}(n\log(n))$ for each detection. Put differently, the time required for the offloading decision would accumulate to $\sum_{j=1}^L \mathcal{O}(n_j\log(n_j))$ when utilizing information from all nearest servers, where $n_{j}$ represents the number of task information provided by server $S_j \in \mathcal{S}_k$.

For each event, we underscore the following key points. Firstly, the information provided to the user comprises an optimal set of tasks in the server's queue. This facilitates the scheduling run for both the server information and the new arrival task, ensuring that optimal schemes output a maximum of one outage. Secondly, in accordance with the procedure outlined in \autoref{alg:1}, when task $T_a$ is under consideration for scheduling, all tasks with cycle requirements smaller than $T_a$ are included in the scheduled list $\mathbf{q}$. Conversely, no tasks with cycle requirements larger than $T_a$ are present in $\mathbf{q}$. Given that \autoref{alg:1} traverses the set of requirement cycles in increasing order, \autoref{alg:3} can be viewed as a derivative of \autoref{alg:1}. Specifically, in its assessment of the outage for a new arrival task, \autoref{alg:3} includes all tasks with smaller or equal requirement cycles to calculate the delay of $T_a$, while disregarding others. Hence, we can ensure the correctness of \autoref{alg:3}.
As \autoref{alg:3} traverses all elements in a sole time,  the worst-case time complexity of the FOD algorithm is $\mathcal{O}(N)$.

\section{Simulation Results}
\label{sec:experiment}
\subsection{Simulation Setup and Parameters}
We now provide numerical results to demonstrate the effectiveness of the proposed algorithms. The simulation is conducted in an area of approximately $5000\times5000$ $[m^2]$, which consists of 20 servers and 30-80 users. We consider that each user has at least two nearest servers less than $1000m$. The other main parameters are given in \autoref{table:2}.
\begin{table}[t]
\caption{System parameters}
\label{table:2}
\begin{tabularx}{0.99\columnwidth}{@{} l |*{3}{C} c @{}}
\toprule
No. & Parameter & Value & Unit\\ 
\midrule
1 & Number of users & 30-80 & \\
2 & Number of servers & 20\\
3 & Arrival rate & 3 & [Tasks/user/second]\\
4 & Average required CPU cycles & 200M & [cycles]\\
5 & CPU capacity in servers & 10-20 & [GHz]\\
6 &  Total system bandwidth & 20 & [MHz]\\
7 & Number of channels & 10 &\\
8 & Data size & 10-5000 &[Kbytes]\\
9 & Path-loss exponent & 3 & \\
10 & Transmission power& 100 & [mW] \\
\bottomrule
\end{tabularx}
\end{table}
For the online schemes, we consider that the arrival event is randomly generated with an exponential distribution with $\lambda = 3$ [tasks/second/user]. In a dynamic environment, users are moved at a constant speed of 10 m/s.

For comparison purposes, we consider Dedas \cite{meng2019dedas} as a benchmark scheme. In addition, the following existing schemes are also considered for performance comparison.
\begin{itemize}
    \item Earlier Deadline First (EDF): EDF is considered the most well-known for the deterministic options. It selects the task for execution based on its deadline, \textit{i.e.} from the smallest to the largest one \cite{zhang2010service}.
    \item Smallest Data Size First (SDF): The task with the smallest data size is selected first \cite{fang2019job}.
    \item The smallest size of $\text{D*S}$: This scheme first calculates the value of $\text{D*S}$ (deadline$\times$data size) for each task in the queue, and then selects the smallest size first for execution \cite{zhang2010service}.
    \item Moore-Hogson (Moore): This scheme selects the task from the smallest deadline, and when an outage occurs, the highest data size will be removed \cite{CHERIYAN2021842}. 
\end{itemize}

We note that Dedas cannot be used in the case of having the same arrival time for all tasks and each has a different deadline. The reason is that it does not have any priority for what kind of task should be selected first.  In contrast, other schemes such as EDF, SDF, and $\text{D*S}$ 
are queuing-based methods that belong to the priority queuing group. Each scheme prioritizes tasks based on different criteria: deadlines, data size, or a combination of data sizes and deadlines.

All the simulation results  are conducted by Python version 3.11.4 under  Windows 11 pro-64-bit (10.0, build 22621); processor: 11th Gen Intel(R) Core(TM) i9-11900K @3.50GHz (16 CPUs), ~3.5 GH; memory: 65534Mb. 

\subsection{Performance Metrics}
To evaluate the performance of the proposed algorithms, we adopt the following two main evaluation metrics:
\subsubsection{Service Ratio}
The service ratio is measured by the number of served tasks over the total tasks in the server within a period. This implies that the higher the service ratio, the better the performance of the considered scheme can be achieved. 

If $N_o$ is the total number of  outages in all the arrival events, $G^{e}$, the service ratio can be calculated as:
\begin{align}
    R_{st} = \frac{G^{e} - N_o}{G^{e}}.
\end{align}

\subsubsection{Running Time}
To measure the scheduling cost, we run different schemes under the same dataset. In particular, to count the scheduling time at each arrival event, we use the function perf\_counter' in package `time', which is provided in the python3 library, to measure how long the scheduling scheme is used for that arrival.

\subsection{Results and Discussion}
\label{sec:results_discussion}
\subsubsection{Service Ratio}

\begin{figure}[t]
\centering
\includegraphics[width=0.95\columnwidth]{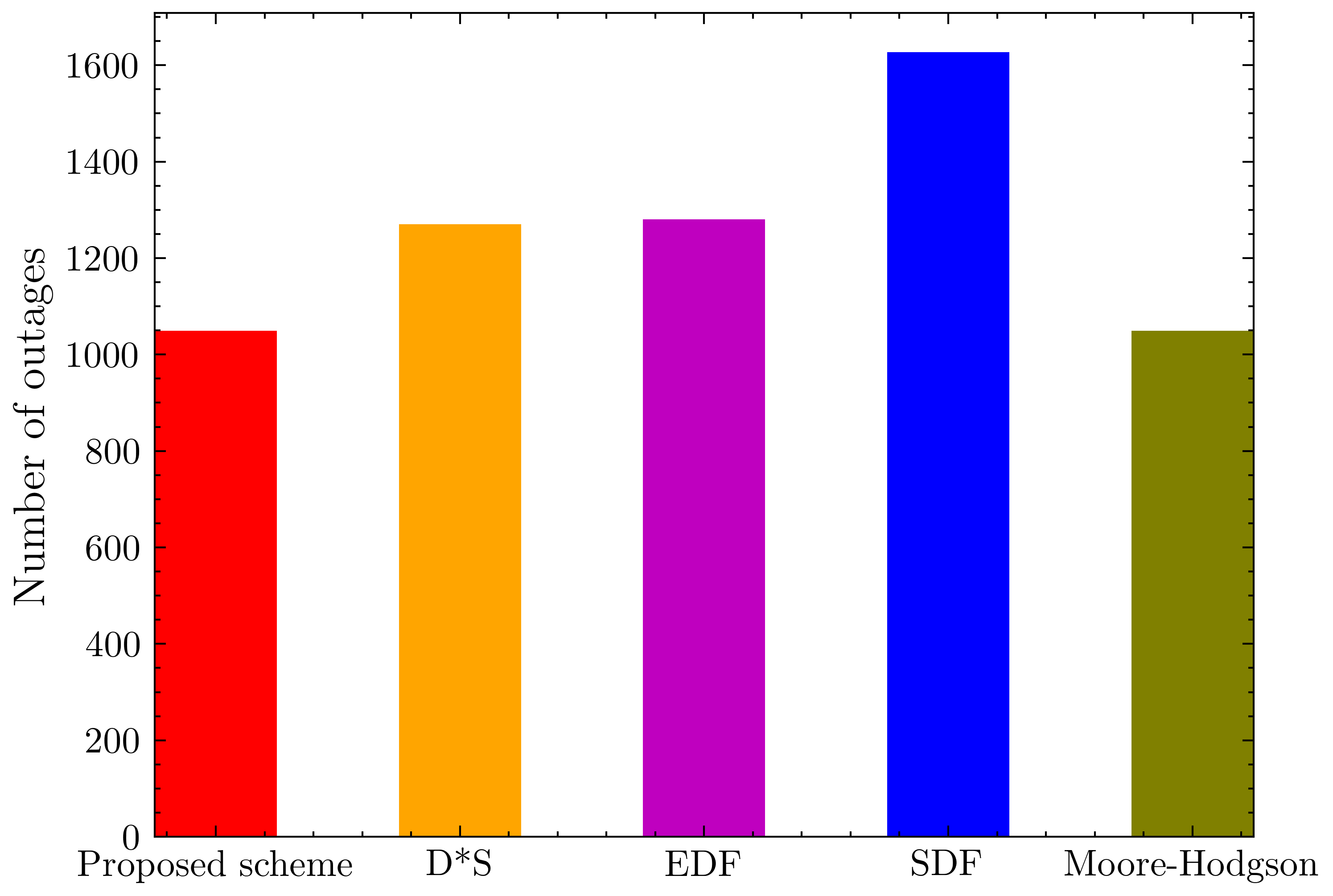}
\caption{Number of outages with different schemes ($f = 4$ [GHz]).}
\label{fig:num_of_outages_4}
\hfill
\end{figure}

\begin{figure}[t]
\centering
\includegraphics[width=0.9\columnwidth]{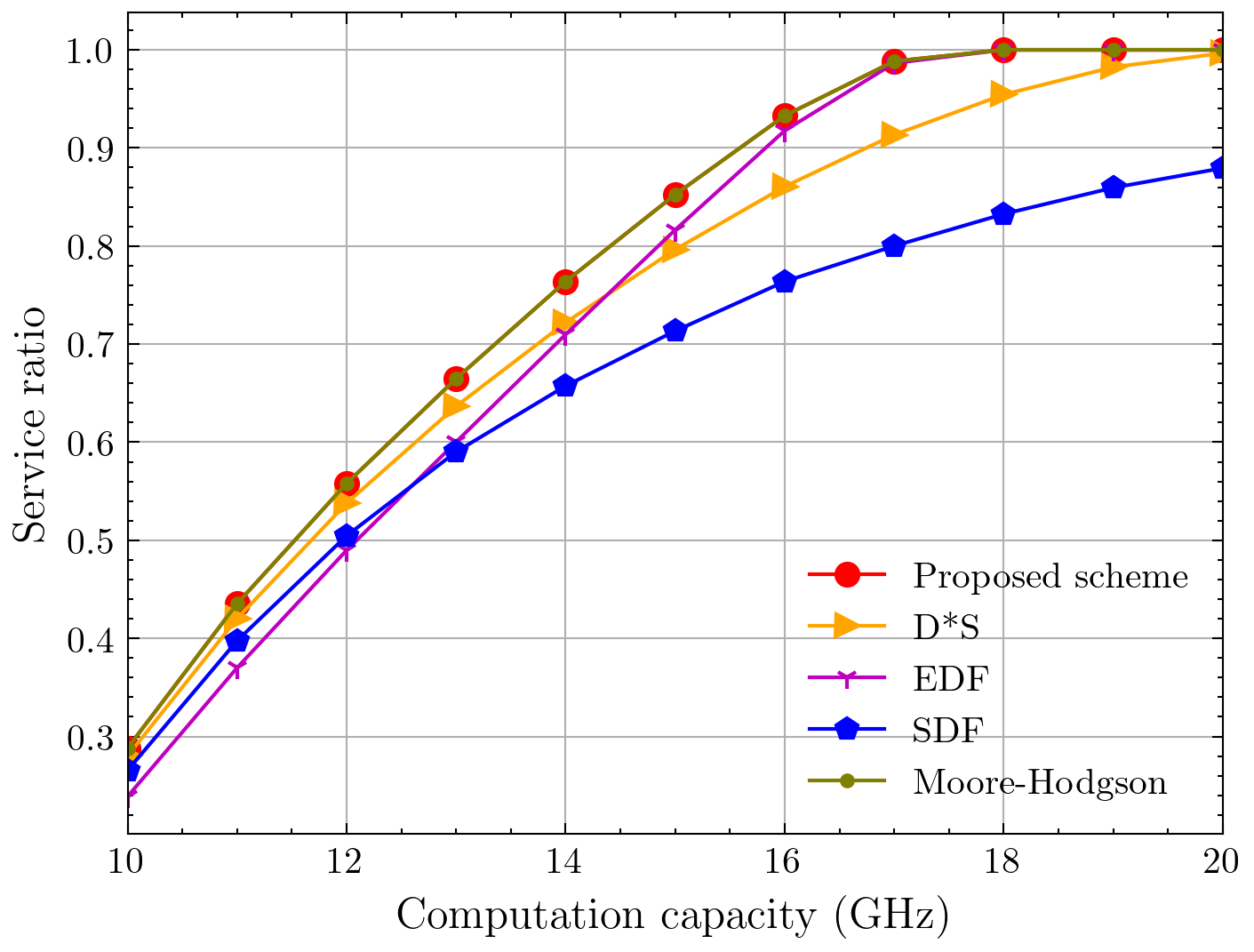}
\caption{Service ratio with different schemes.}
\label{fig:service_ratio}
\hfill
\end{figure}

As illustrated in \autoref{fig:num_of_outages_4}, we run the considered schemes under the assumption that all tasks have the same arrival time. The number of outages is counted over 10000 tasks. As seen in the figure, the proposed algorithm and Moore-Hodgson provide the best performance in terms of the number of outages (e.g. 1010). The other schemes perform much worse compared to the proposed scheme (e.g. the number of outages is in the range of 1200-1600).  In contrast to alternative schemes, the proposed approach employs a dual selection criterion for tasks. While SDF prioritizes tasks solely based on their required CPU cycles, the $\text{D*S}$ scheme gives precedence to tasks using a calculation involving their required CPU cycles, deadline, and current clock value. On the other hand, EDF organizes tasks based solely on their deadlines.

 \autoref{fig:service_ratio} plots the service ratio with different schemes, where the different CPU capacities are considered. As we can observe for a small computation capacity the service ratio of all schemes is low and the gap among these schemes is quite negligible. As expected at high computation capacity, the service ratio of all schemes is largely improved. Again, the proposed scheme and Moore-Hodgson provide the best performance. Interestingly, EDF obtains a similar service ratio when the computing capacity is larger than 18 [GHz], while SDF is worse. 

\begin{figure}[t]
\centering
\includegraphics[width=0.9\columnwidth]{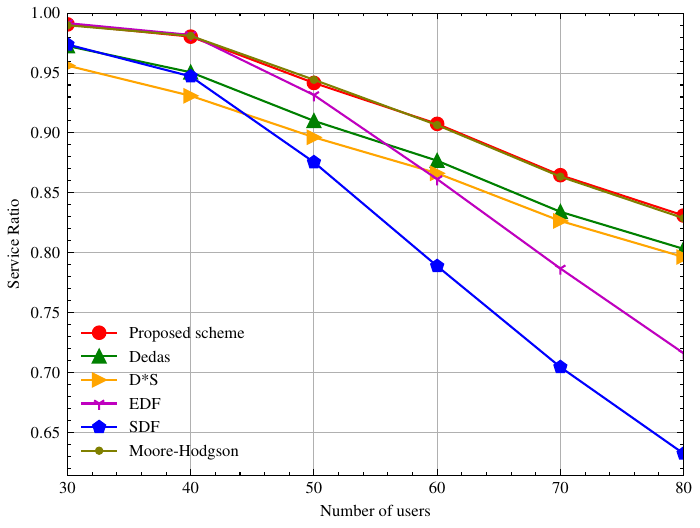}
\caption{Service ratio with different numbers of users.}
\label{fig:online_sr_u}
\hfill
\end{figure}

\begin{figure}[t]
\centering
\includegraphics[width=0.9\columnwidth]{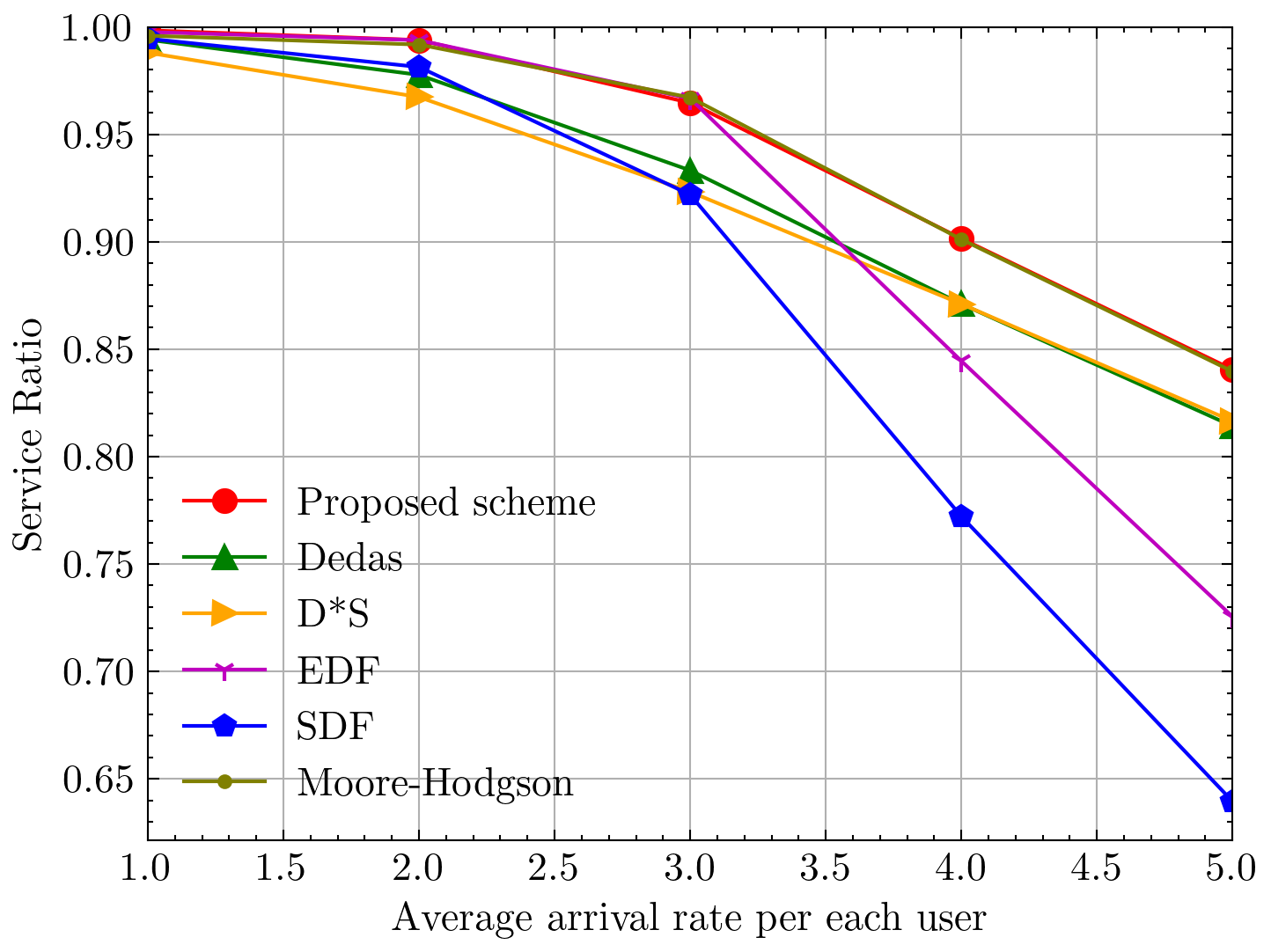}
\caption{Service ratio with different rates.}
\label{fig:c_rate}
\hfill
\end{figure}

\begin{figure}[t]
\centering
\includegraphics[width=0.9\columnwidth]{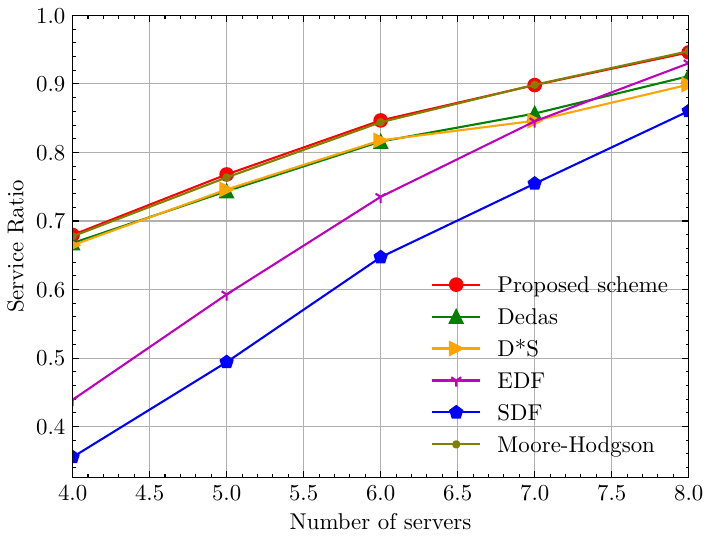}
\caption{Service ratio with different numbers of nearest servers.}
\label{fig:online_sr_s}
\hfill
\end{figure}

  We now show the service ratio for online schemes
 in  Figs. \ref{fig:online_sr_u}, \ref{fig:c_rate} and \ref{fig:online_sr_s}. Particularly, in \autoref{fig:online_sr_u}, the number of users in the system is shown in the range of $[30:80]$. In  \autoref{fig:c_rate}, we fix the number of users in the systems at 50 and change the arrival rates in increments from 1 to 5. \autoref{fig:online_sr_s} shows the service ratio when the number of users is fixed at 50 and the number of nearest servers is changed from 4 to 8. It is essential to highlight that, in this study, the challenge of task scheduling arises due to users' lack of awareness regarding the number of tasks that may precede their offloading task. Consequently, the scheduling problem becomes inherently uncertain. Again, the proposed scheduling scheme and Moore-Hodgson offer the highest ratio of successfully arranged tasks. This phenomenon demonstrates the effectiveness of the proposed scheme in addressing the uncertainty problem due to the lack of awareness of the arrival tasks.
 
\subsubsection{Running time}
\begin{figure}[t]
\centering
\includegraphics[width=0.9\columnwidth]{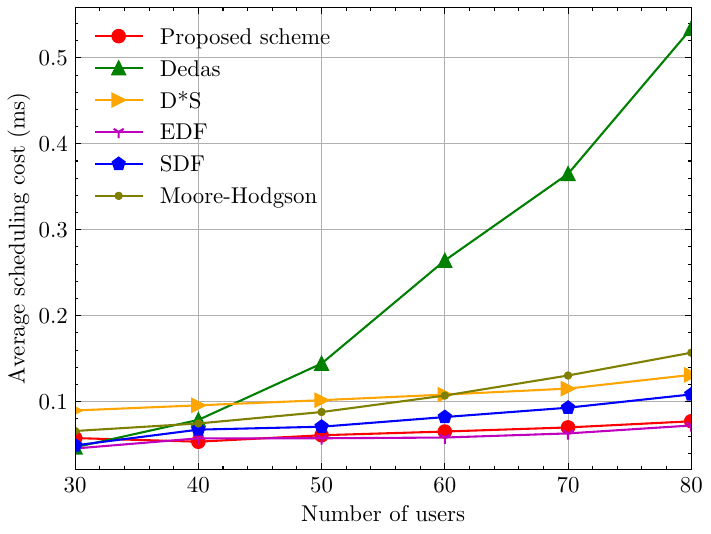}
\caption{Average scheduling cost (seconds/task) with different numbers of users.}
\label{fig:schedulingcost_us}
\hfill
\end{figure}

We now evaluate the performance of the proposed scheme in terms of running time, \textit{i.e.} scheduling cost (seconds/task), as illustrated in \autoref{fig:schedulingcost_us}. Notably, when the system accommodates a low number of users, the running cost of scheduling remains relatively low. This can be attributed to the reduced total arrival rate, resulting in fewer tasks offloading to the server, and thus minimal scheduling costs. However, as the number of users increases, particularly in the range of [50:80], the scheduling costs vary significantly. First, we can observe that the proposed scheme exhibits the lowest scheduling cost. This result stems from \autoref{alg:3}, which is used by the user to check task acceptance with all provided servers' information. It is essential to recall that \autoref{alg:3} is a consequence of \autoref{alg:1}. Therefore, the orders provided by this scheme are always valid. It is noted that Dedas and Moor-Hodgson provide the worst performance (\textit{i.e.} the highest cost) followed by $\text{D*S}$, which has a time complexity of $\mathcal{O}(N\log N)$.

Next, the scheme with the second-highest scheduling cost is Dedas, which adopts a strategy involving scanning all positions in the queue for new arrivals. At each position, Dedas assesses the feasibility of accommodating the new task. Despite the anticipated higher scheduling cost resulting from this exhaustive scanning process, Dedas manages to optimize its performance. It initially appends the new arrival as a priority case at the end of the queue. Only if this initial attempt fails.  Dedas proceeds with a full scan of all positions, effectively striking a balance between efficiency and thoroughness.

The proposed algorithm, EDF and SDF offer remarkably low scheduling costs. The efficiency of these schemes can be attributed to their simplicity. In the case of EDF and SDF, the process entails identifying a position that fulfills the priority based on the deadline and data size, respectively. In our proposed scheme, tasks with data sizes exceeding the system's capacity are initially excluded. Subsequently, the new arrival is positioned based on an increasing deadline order, further enhancing the overall efficiency of the scheduling process.

\begin{figure}[t]
\centering
\includegraphics[width=1\columnwidth]{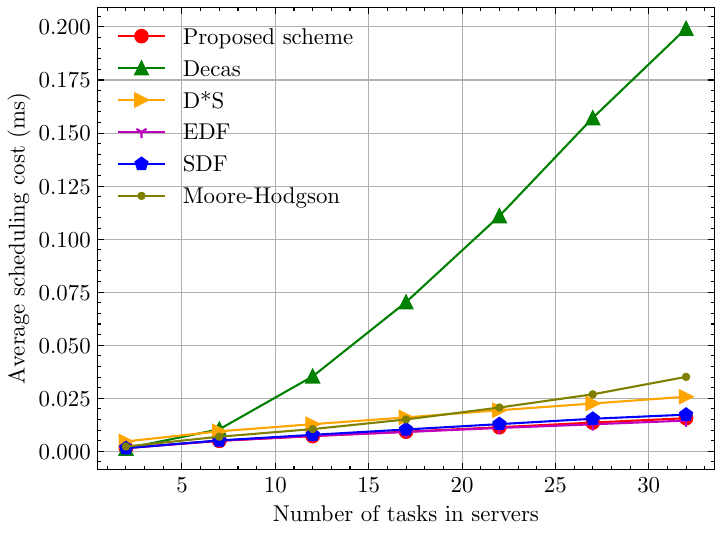}
\caption{Average scheduling cost based on different numbers of tasks.}
\label{fig:schedulingcost_ts}
\hfill
\end{figure}

Lastly, in \autoref{fig:schedulingcost_ts}, we evaluate the scheduling time with respect to the number of tasks in MEC servers. The observed scheduling time closely aligns with theoretical expectations, wherein SDF, EDF and \autoref{alg:3} exhibit linear time complexity $\mathcal{O}(N)$. This similarity to \autoref{fig:schedulingcost_us} further underscores the equivalence between varying the total number of users and adjusting the arrival rate per participant, as illustrated in \autoref{fig:online_sr_u} and \autoref{fig:online_sr_s}.

In a nutshell,  the numerical results demonstrate that the proposed scheme not only performs excellently across the considered metrics but also scales effectively as the number of users in the system increases. When examining service ratios in relation to variations in the number of users, arrival rates, and nearest servers, it is observed that at low system loads, the service ratio remains similar across almost all schemes. However, as the load increases, the service ratios of the compared schemes decrease significantly, whereas the proposed scheme experiences only a slight decline. A similar trend is observed for scheduling costs; when the system load is low and the number of tasks in the server's queue is small, nearly every scheme incurs similar scheduling costs. As the number of users in the system increases, each server's workload also increases. The results for the proposed scheme show the lowest scheduling cost, with minimal variation as the number of users changes, which is a significant advantage compared to other schemes.

\section{Conclusion}
\label{sec:conclus}
This study has focused on the exploration of deadline-aware task scheduling within MEC systems, presenting innovative solutions and insights. In particular, we have presented a novel offloading policy to empower users to directly engage with servers for offloading requests. This solution enables users to make informed decisions based on the information provided by the servers. We have put forward an optimal job scheduling algorithm, supported by a rigorous correctness proof, and an online scheme designed to minimize scheduling costs while maximizing performance. Extensive numerical results have been provided to demonstrate the effectiveness of the proposed solution by showing
a significantly superior performance compared to alternative schemes, in terms of the service ratio and scheduling cost. 

\appendices
\section{Proof of \cref{lem:1}}
\label{p:lamma1}
We start with the case in which $\mathbf{s}$ satisfies $\beta_{s(i)}< \beta_{s(j)}$ for all $i, j$.
Then, from the definition, $\mathbf{s}=\mathbf{s}'$ and $\mathbf{s}'$ has no outage.
Next, we consider the case $\mathbf{s}\neq \mathbf{s}'$, in which there exists an index pair $i$ and $j$ such that $\beta_{s(i)} > \beta_{s(j)}$ with $j = i+1$ and $i, j \in [1, S]$. Here, if $\mathbf{s}$ has no outage, then $\sum_{k=1}^j \frac{\alpha_{s(k)}}{f} = \sum_{l=1}^i \frac{\alpha_{s(l)}}{f} +\frac{\alpha_{s(j)}}{f} \leq \beta_{s(j)} < \beta_{s(i)}$. We can see that the exchanged orders of $s(i)$ and $s(j)$ do not make any outage in the range of $[1:i-1]$ and $[j+1:S]$. If we exchange the orders, then
$\sum_{k=1}^i \frac{\alpha_{s(k)}}{f} = \sum_{l=1}^j \frac{\alpha_{s(l)}}{f} +\frac{\alpha_{s(i)}}{f} \leq \beta_{s(j)} < \beta_{s(i)}$ is satisfied. Therefore, there is no outage after the exchange.
If $\beta_{s(i)} > \beta_{s(j)}$, the index pair ($i,j$) with $j = i+1$ and $i, j \in [1, S]$ still exists in the set $\mathbf{s}$. We keep the above exchanging steps until the condition is fulfilled. Then both sets $\mathbf{s}$  and $\mathbf{s}'$ are identical.

\section{Proof of \cref{lem:3}}
\label{p:lamma3}
In this proof, let $\mathbf{s}$ be an ordered set of the increasing deadlines in $[i:j]$. We can show that $s(k)$ belongs to $\mathbf{s}'$ for any index $k\in [i:j]$ such that $\alpha_{s(k)} < \alpha_{s'(k)}$, .
     Assume that $s(k)$ is the first task satisfying $\alpha_{s(k)} < \alpha_{s'(k)}$ in $[i:j]$. Since the condition $\sum_{l=1}^{k} \alpha_{s(l)}\leq \beta_{s(k)}$ holds true in $\mathbf{s}$, if $s(k)$ is in the set $\mathbf{s}'$ and $s'(k)$ is not in $\mathbf{s}'$ with $[i+1:k-1]$, there is no  outage, and $s(k)$ is selected before $s'(k)$. As a result, if $s'(k)$ is in $\mathbf{s}'$, $s(k)$ is also in $\mathbf{s}'$. Due to the deadline constraints, we notice that $\beta_{s(k)} \leq \beta_{s'(k)}$, and therefore $s(k)$ should be placed before $k$ in  $[i:j]$.

    Similarly, for the second element satisfied $\alpha_{s(k)} < \alpha_{s'(k)}$, we prove that $s(k)$  also belongs to the set $\mathbf{s}$. Since each element $a\in [i:k-1]$ with $\alpha_{s(a)} < \alpha_{s'(a)}$, $s(a)$ also belongs to $\mathbf{s}'$. Therefore, it is true that $\sum_{l=1}^k \alpha_{s'(l)} \leq \sum_{l=1}^k \alpha_{s(l)} \leq \beta_{s(s)}$. This also implies that if $s(k)\in\mathbf{s}'$ and $s'(k) \notin \mathbf{s}'$, there is no outage in  $\mathbf{s}'$. Because the deadline constraint $\beta_{s(k)} \leq \beta_{s'(k)}$, Task $s(k)$ only appears in the range  $[i:k-1]$. 
    In summary, we can conclude $\sum_{l=i}^j\alpha_{s'(l)} \leq \sum_{l=i}^j\alpha_{s(l)}$.
\section{Proof of \cref{lem:4}}
\label{p:lamma4}

\begin{algorithm}[t]
\caption{Replacing Steps for case 2 ($\alpha_{s'(i)} \leq \alpha_{s(i)}$ and $\beta_{s'(i)} > \beta_{s(i)}$)}
\label{alg:2}
\begin{algorithmic}[1]
\State{{\bf Input}: The ordered set  $\mathbf{s}' =  [s'(1), s'(2), \cdots, s'(i-1), s'(i), \cdots,  s'(P)]$, the optimal ordered set $\mathbf{s} =  [s'(1), s'(2), \cdots, s'(i-1), s(i), \cdots,  s(M)]$ and the current replacing index $i$.}

\State{{\bf Initialization}: $m\gets 0$, $l\gets 0$, and $p\gets 0$; }
\If{$s'(i) \notin \mathbf{s}$}
    \State{Update $\mathbf{s}$ with element-wise: $s(i) \gets s'(i)$. }
    \State{$m \gets i$.}
\ElsIf{ $s'(i) = s(j) \in \mathbf{s}$ with $j\in [i+1:M]$}
    
    \While{$s'(k)=s(j')$ with $j'\in [j+1:M]$, $k\in [i+1:j]$}
        \State{$j \gets j'$.}
    \EndWhile
    \State{$m \gets \min(j, P)$.}
    \State{Update $\mathbf{s}$ with block-wise: $s([i:m]) \gets s'([i:m])$}
\EndIf

\State{Find index $p$ of elements in set $\mathbf{s}$ as \eqref{eq:4}}
\begin{align}
    & \label{eq:3} \mathbf{l} = [b \in [m:P]| \beta_{s'(b)} > \beta_{s(b+1)}] \\
    & \label{eq:4} p = 
    \begin{cases}
         a+1; a=\min{(\max{(\mathbf{l})}, P)}.\\
        -1; \mathbf{l} = \text{\o}.
    \end{cases}
\end{align}
\If {$p \neq -1$}        
    \State{Update $\mathbf{s}$ with block-wise: $s([m+1:p]) \gets s'([m+1:p])$}
\EndIf
\end{algorithmic}
\end{algorithm}

To prove \cref{lem:4}, we assume that all elements from $[1:i-1]$ for any $i \in [1:M]$ in $\mathbf{s}$ are the same as $\mathbf{s}'$. Therefore, the proposed algorithm produces the ordered set $\mathbf{s}' =[s'(1), s'(2), \cdots, s'(i-1), s'(i),\cdots ,s'(P)]$, while the ordered set $\mathbf{s} =[s'(1), s'(2), \cdots, s'(i-1), s(i),\cdots ,s(M)]$ is an optimal set that satisfies that $\beta_{s(i)} < \beta_{s(j)}$ for any $i, j \in [1:M]$ and $i<j$. It is clear that the size of $\mathbf{s}$ is greater than or equal to that of $\mathbf{s}'$ or $M\geq P$. By replacing $s(i)$ in $\mathbf{s}$ with $s'(i)$ in $\mathbf{s}$, we first prove that the replacement doesn't cause any outages in optimal ordered set $\mathbf{s}$. The replacement operation can happen in four cases:
    \begin{itemize}
        \item \textit{\textbf{Case 1}:} $\alpha_{s(i)} \geq \alpha_{s'(i)}$ and  $\beta_{s(i)} \geq  \beta_{s'(i)}$. We first note that $s'(i)$ satisfies the deadline requirement in set $\mathbf{s}'$ that is $\sum_{k=1}^{i}\frac{\alpha_{s'(k)}}{f} \leq \beta_{s'(i)}$; and  all elements in $\mathbf{s}$ are the same until $i-1$. It means when replacing $s(i)$ by $s'(i)$, there is no outage for task $T_{s'(i)}$. Next, the condition $\alpha_{s'(i)}<\alpha_{s(i)}$  means that when replacing $s(i)$ by $s'(i)$, the delay $t_{s(k)}$ of task $T_{s(j)}$ is reduced an amount equal to $\frac{\alpha_{s(i)}-\alpha_{s'(i)}}{f}$. In other words, replacing $s(i)$ with $s'(i)$ doesn't cause any outages for any tasks $T_{s(k)}$ with $k\in [1:M]$ and $s(k)\in \mathbf{s}$. Under the increasing deadlines and $\beta_{s(i)} \geq  \beta_{s'(i)}$,  there will be no order $s'(i)$ of task $T_{s'(i)}$ appear behind the order $s(i)$ of task $T_{s(i)}$. Consequently, if we replace order $s(i)$ by $s'(i)$, the set $\mathbf{s}$ is valid with all elements to be unique.
        
        \item \textit{\textbf{Case 2}:} $\alpha_{s(i)} \geq \alpha_{s'(i)}$ and $\beta_{s(i)} < \beta_{s'(i)}$.
        In this case, with the element-wise (\textit{\textbf{Step 2.1}}) or block-wise replacement (\textit{\textbf{Step 2.2}}), we prove that it doesn't make any outages and duplicates. After replacement, we check the updated set to verify it is an increasing set of deadlines (\textit{\textbf{Step 2.3}}). The detail is given in \autoref{alg:2}.
        \begin{itemize}
            \item 
        \textit{\textbf{Step 2.1}:} If the index $j$ with $j\in [i+1:P]$ does not exist such that $s'(i) = s(j)$, the element-wise replacement $s(i)$ with $s'(i)$ will not make any duplicate in set $\mathbf{s}$. As $\alpha_{s(i)} \geq \alpha_{s'(i)}$, the delay of each task after $i$ in $\mathbf{s}$ is reduced by the amount equal to $\frac{\alpha_{s(i)}- \alpha_{s'(i)}}{f}$. Thus, the replacement does not cause any outage. It is noted that we set $m=i$ for the last replaced element.

        \item \textit{\textbf{Step 2.2}:} If there is an index $j \in [i+1:P]$ such that element $s'(i) = s(j)$ with block-wise replacement in the range of $[i:j]$, there is no duplicate after replacement.

        As shown in \autoref{alg:2}, we have $\beta_{s(m)} \leq \beta_{s'(m)}$ with $m$ being the last replaced element. This is because there   always exists at least an index $l \in [i:m-1]$ in the range of $[i:m-1]$ such that $s'(l) = s(j)$ with $j\in [m:M]$ and $\beta_{s(m)}\leq \beta_{s(j)}=\beta_{s'(l)} \leq \beta_{s'(m)}$. Combining with the conditions $\alpha_{s'(i)} \leq \alpha_{s(i)}$, $\beta_{s'(i)} \geq \beta_{s(i)}$ and \cref{lem:3}, we have $\sum_{l=i}^m\alpha_{s'(l)} \leq \sum_{l=i}^m\alpha_{s(l)}$. That implies that the block-wise replacement does not cause any outages for tasks $j'$ with $j'>j$.

        After \textit{\textbf{Steps 2.1}} and \textit{\textbf{2.2}}, we need to find the index $p$ in \eqref{eq:4}. If the set $\mathbf{s}$ is an increasing set of deadlines, we set  $p=-1$ as a default value.

         \item \textit{\textbf{Step 2.3}:}    $p \neq -1$ means that the set $\mathbf{s}$  is not an increasing set of deadlines after updating at \textit{\textbf{Step 2.1}} or \textit{\textbf{Step 2.2}}.
                 From \eqref{eq:3}, \eqref{eq:4} and $\beta_{s(l)}<\beta_{s'(l-1)} \leq \beta_{s'(l)}$, we have $\beta_{s'(l)} > \beta_{s(l)}$ for any $l\in [m+1:p]$. We also have $\beta_{s(m+1)} < \beta_{s'(m+1)}$ and tasks from 1 to $m$ will be identical in two sets $\mathbf{s}$ and $\mathbf{s}'$. Assuming $\alpha_{s(m+1)} < \alpha_{s'(m+1)}$, we have $\sum_{k=1}^{m+1} \alpha_{s(k)} < \sum_{k=1}^{m+1} \alpha_{s'(k)}$. Here, $\alpha_{s(m+1)}$ is selected before $\alpha_{s(m+1)}$. In other words, $s(m+1)$ should be scheduled in $\mathbf{s}'$. That conflicts with the condition that tasks from 1 to $m$ are the same in two sets $\mathbf{s}$ and $\mathbf{s}'$. Thus,  $\alpha_{s(m+1)} \geq \alpha_{s'(m+1)}$ holds true.
                 
         As a result, we have $\sum_{l=m+1}^p \alpha_{s'(l)} \leq \sum_{l=m+1}^p \alpha_{s(l)}$, and the block-wise replacement of tasks in $[m+1:p]$ does cause any outage for tasks after $p$.
         Since $\beta_{s'(p)}\leq \beta_{s(p+1)}$, there is no duplicate in set $\mathbf{s}$ after the block-wise replacement, making the set $\mathbf{s}$ to be an increasing set of deadlines.       
        \end{itemize}
    
        \item \textit{\textbf{Case 3}:} $\alpha_{s(i)} < \alpha_{s'(i)}$ and $\beta_{s(i)} \leq \beta_{s'(i)}$. Under the deadlines' requirements and the amounts of required CPU cycles between two tasks $T_{s(i)}$ and $T_{s'(i)}$, the proposed algorithm will place $s(i)$ in front of $s'(i)$ due to $\alpha_{s(i)}< \alpha_{s'(i)}$. However, it is true that $\beta_{s'(i-1)} = \beta_{s(i-1)}$, $\beta_{s'(i-1)} \leq \beta_{s'(i)}$ and $\beta_{s(i-1)} \leq \beta_{s(i)}$, and thus the condition $\sum_{j=1}^{i} \frac{\alpha_{s'(j)}}{f} \leq \beta_{s(i)} \leq \beta_{s'(i)}$ is met. The delay $t_{s'(j)}$ of task $T_{s'(j)}$ is not an outage with any $j \in [i+1:N]$ due to $\alpha_{s(i)} < \alpha_{s'(i)}$. It means there is no outage if considering and ordering $s(i)$ first and in front of $s'(i)$. Therefore, this case cannot happen because of conflict with the assumption.
        
        \item \textit{\textbf{Case 4}:} $\alpha_{s(i)} < \alpha_{s'(i)}$ and $\beta_{s(i)} > \beta_{s'(i)}$. Since this case satisfies the condition of \autoref{lem:2}, then $s(i)$ belongs to $\mathbf{s}'$. By denoting $s'(j) = s(i)$ in $\mathbf{s}$ with $j \in [i+1:P]$, we have $\sum_{k=i}^{j} \frac{\alpha_{s'(k)}}{f}\leq \sum_{k=i}^{j} \frac{\alpha_{s(k)}}{f}$. If $\sum_{k=i}^{j} \frac{\alpha_{s'(k)}}{f} > \sum_{k=i}^{j} \frac{\alpha_{s(k)}}{f}$, there is at least a task $T_{s'(l)}$ with $l \in [i+1:j]$ such that $\alpha_{s'(l)} > \alpha_{s(l)}$. We note that both sets $\mathbf{s}$ and $\mathbf{s}'$ satisfy $\beta_{s(m)} < \beta_{s(n)}$ and $\beta_{s'(m)} < \beta_{s'(n)}$ for all $m, n \in [1:M]$ or $m, n \in [1:P]$. Furthermore, we have $\beta_{s(i)}$ is smallest in the range $[i+1:j]$, and $\beta_{s(i)} = \beta_{s'(j)}$ is largest in the range $[i+1:j]$. Therefore, two elements $s'(l)$ and $s(l)$  satisfy the constraint of \cref{lem:2}. Thus, there exists an index $k \in [l+1:P]$ such that $s(l) = s'(k)$. \autoref{alg:1} finds the element $s(k)$ to meet the condition $\sum_{m=i}^{k} \frac{\alpha_{s'(m)}}{f}\leq \sum_{m=i}^{k} \frac{\alpha_{s(m)}}{f}$. Therefore, we can replace the whole block of element $[s(i):s(j)]$ or $[s(i):s(k)]$ by $[s'(i):s'(j)]$ or $[s'(i):s'(k)]$ without outages with $i \in [1:P]$, $j \in [i+1:P]$, $l\in [i+1:j]$ and $k\in [l+1:P]$.
        \end{itemize}
        
    By the above cases, we have shown that there is no outage if we replace elements from $\mathbf{s}'$ to $\mathbf{s}$. Secondly, we prove that after replacing $P$ elements in $\mathbf{s}$ with $\mathbf{s}'$, the remaining of the ordered set $\mathbf{s}$ in $[P+1:M]$ is empty. Thus, the subset of the remaining ordered set is denoted by  $\mathbf{s}_{r} = [s_{r}(P+1), s_{r}(P+2), \cdots, s_{r}(M)]$, where $\beta_{s'(i)} \leq \beta_{s_{r}(j)}$ with $i \in [1:P]$ and $j \in [P+1:M]$. If task $T_{s_{r}(j)}$ with $j\in [P+1:M]$ satisfies $\alpha_{s_{r}(j)} < \alpha_{s'(i)}$ with $i \in [1:P]$, $s_r(j)$ should be in the set $\mathbf{s}$ (as in \cref{lem:2}). There will be no task $T_{s_{r}(j)}$ in $\mathbf{s}_{r}$ with $j\in [P+1:M]$ satisfying $\alpha_{s_{r}(j)} < \alpha_{s'(i)}$. Next, \autoref{alg:1} finds task $T_{s_{r}(j)}$ with $j\in [P+1:M]$ and adds it in the $\mathbf{s}'$ if $\alpha_{s_{r}(j)} > \alpha_{s'(i)}$.
   Since this conflicts with the stopping condition of \autoref{alg:1}, the set $\mathbf{s}_{r}$ is empty or $P=M$. The proof of \cref{lem:4} is thus complete.

\bibliographystyle{./trans/IEEEtran}
\bibliography{ ./trans/IEEEabrv, refs}
% \printbibliography

\end{document}